\documentclass[
 reprint,
nofootinbib,
 amsmath,amssymb,
 aps,
]{revtex4-2}
\usepackage{graphicx}
\usepackage{hyperref}
\usepackage{dcolumn}
\usepackage{bm}
\usepackage[mathlines]{lineno}
\usepackage{verbatim}
\begin{document}
\preprint{APS/123-QED}
\title{Role of individual components of two-nucleon interaction in nuclear matrix elements of $2\nu\beta\beta$ and $0\nu\beta\beta$ of $^\textbf{48}$Ca: Beyond the closure approximation}
\author{Shahariar Sarkar}
\email{shahariar.sarkar@iitrpr.ac.in}
\affiliation{Indian Institute of Technology Ropar, Rupnagar, Punjab-140001, India}
\date{\today}
\author{Pawan Kumar}
\affiliation{Indian Institute of Technology Ropar, Rupnagar, Punjab-140001, India}
\date{\today}
\author{Kanhaiya Jha}
\affiliation{Indian Institute of Technology Ropar, Rupnagar, Punjab-140001, India}
\date{\today}
\author{P.K. Raina}
\affiliation{Indian Institute of Technology Ropar, Rupnagar, Punjab-140001, India}
\date{\today}
\begin{abstract}
In the present work, we examine the role of central (C), spin-orbit (SO) and tensor (T) components of two-nucleon interaction in the nuclear matrix elements (NMEs) of the two-neutrino double beta decay ($2\nu\beta\beta$) and the light neutrino-exchange mechanism of neutrinoless double beta decay ($0\nu\beta\beta$) of $^{48}$Ca in closure approximation and nonclosure approach. The NMEs are calculated in the nuclear shell-model framework using two-nucleon effective interaction GXPF1A used for $pf$ shell. The decomposition of the shell model two-nucleon interaction into its individual components is performed using the spin-tensor decomposition (STD). The NMEs for $2\nu\beta\beta$ are calculated in running nonclosure method. The NMEs for $0\nu\beta\beta$ are calculated with four different methods ,namely, closure, running closure, running nonclosure, and mixed method.
Results show that the magnitude of NMEs for $2\nu\beta\beta$ decreases about 7\% with the C+SO component of the interaction as compared to the C component. The magnitude of NMEs is further decreased about 9\% by adding T component to the C+SO component. The NMEs of $0\nu\beta\beta$ calculated in running nonclosure method are enhanced by about 8-10\%, 8-10\%, and 9-12\%, respectively, as compared to corresponding NMEs calculated in running closure method with C, C+SO components and total (C+SO+T) GXPF1A interaction for different SRC parametrization.
For both $2\nu\beta\beta$ and $0\nu\beta\beta$, the NMEs calculated with C+SO component is in opposite phase with the NMEs calculated with C component and the total GXPF1A interaction. 
\end{abstract}
\maketitle
\section{\label{sec:level1}Introduction}
Two-neutrino double beta decay $(2\nu\beta\beta)$ is a rare second-order weak nuclear decay in which two neutrons inside
an even-even nucleus are simultaneously transformed into two protons accompanied by the emission of two
electrons and two antineutrinos. The process was first suggested by Goeppert Mayer \cite{PhysRev.48.512} in 1935. Almost 80 years
later, $2\nu\beta\beta$ has been observed in 12 nuclei with half-lives ranging from $\sim 10^{19}$ to $10^{24}$ years \cite{saakyan2013two}. Neutrinoless double beta decay ($0\nu\beta\beta$) is another mode of double beta decay in which neutrino comes in the virtual intermediate state; thus, it violates lepton number by two units. In 1939, Wolfgang Furry discussed $0\nu\beta\beta$ for the first time  \cite{PhysRev.56.1184,vergados2012theory} based on E.Majorana's  symmetric theory for fermion and anti-fermion \cite{majorana1937teoria} followed by G. Racah's chain reactions \cite{racah1937sulla} in 1937. While $0\nu\beta\beta$ is allowed theoretically for several nuclei, this process is still unobserved even after 80 years of its prediction. If this process is observed, one can conclude that neutrinos are Majorana Fermion rather than Dirac Fermion \cite{PhysRevD.25.2951}, which has important implications in physics beyond the standard model \cite{deppisch2012neutrinoless,PhysRevD.25.2951,rodejohann2011neutrino}. Also, it gives some hints about neutrino mass and neutrino mass hierarchy \cite{RevModPhys.80.481,vergados2012theory}. 

The half-life of $2\nu\beta\beta$ and $0\nu\beta\beta$ are related with nuclear matrix elements (NMEs) which are calculated theoretically with different many-body nuclear models \cite{engel2017status} such as quasiparticle random phase approximation \cite{vergados2012theory}, interacting nuclear shell-model \cite{PhysRevLett.100.052503,PhysRevC.81.024321,PhysRevC.88.064312,PhysRevLett.113.262501,PhysRevLett.116.112502}, interacting boson model \cite{PhysRevC.79.044301,PhysRevLett.109.042501}, generator coordinate method \cite{PhysRevLett.105.252503}, energy density functional theory \cite{PhysRevLett.105.252503,PhysRevC.90.054309} and the projected Hartree-Fock Bogolibov model \cite{PhysRevC.82.064310}.
 For $0\nu\beta\beta$, various decay mechanisms such as light neutrino exchange mechanism \cite{rodin2006assessment,PhysRevC.60.055502}, heavy neutrino exchange mechanism \cite{vergados2012theory}, left-right symmetric mechanism \cite{PhysRevLett.44.912,PhysRevLett.47.1713}, and supersymmetric particles exchange mechanism \cite{PhysRevD.34.3457,vergados1987neutrinoless} have been proposed. Here, our interest is light neutrino-exchange mechanism.

In the present work, NMEs for $2\nu\beta\beta$ and the light neutrino-exchange mechanism of $0\nu\beta\beta$ are calculated for $^{48}$Ca using interacting nuclear shell-model. The $2\nu\beta\beta$ process for $^{48}$Ca is written as
\begin{equation}
    ^{48}\text{Ca}\rightarrow^{48}\text{Ti}+e^-+e^-+\overline {\nu}_e+\overline {\nu}_e.
\end{equation} The $0\nu\beta\beta$ process for $^{48}$Ca is written as
\begin{equation}
    ^{48}\text{Ca}\rightarrow^{48}\text{Ti}+e^-+e^-.
\end{equation}
The NMEs for $2\nu\beta\beta$ were calculated for $^{48}$Ca in Refs  \cite{PhysRevC.75.034303,PhysRevC.42.1120,caurier1990full} using interacting shell-model. In Refs. \cite{PhysRevC.81.024321,PhysRevLett.113.262501,PhysRevLett.116.112502}, NMEs for the light neutrino-exchange mechanism of $^{48}$Ca were calculated using the closure approximation within the nuclear shell-model. The nonclosure approach was used to calculate NMEs of $0\nu\beta\beta$ of $^{48}$Ca in Ref. \cite{PhysRevC.88.064312} using nuclear shell model. All the above studies have been performed using total effective interaction. However, in the present study we are examining the
contribution of individual components, i.e., central (C), spin-orbit (SO) and tensor force (T), of shell model two-nucleon effective interaction in NMEs of $2\nu\beta\beta$ and $0\nu\beta\beta$. 

In the last few years, the contribution of individual components of two-nucleon interaction gains a lot of interest in understanding the cause of shell evolution in neutron-rich nuclei \cite{kumar2019quasi,PhysRevC.100.024328,PhysRevC.69.024306,PhysRevC.74.034330,PhysRevC.86.034314}. The sensitivity of NMEs for the light neutrino-exchange mechanism of $0\nu\beta\beta$ of $^{48}$Ca with individual components of two-nucleon interaction was also studied recently in Ref. \cite{PhysRevC.101.014307} using closure approximation. In the present study, we go beyond the closure approximation to examine the role of individual components of two nucleon interaction in NMEs of $2\nu\beta\beta$ and $0\nu\beta\beta$. The decomposition of effective shell model interaction into its C, SO and T force components is performed using STD \cite{dirim1968jp,Kirson:1973ffz,RevModPhys.48.191,PhysRevC.15.1483,kenji1980spin,smirnova2010shell,brown1985spin,PhysRevC.45.662,PhysRevC.86.034314}. The STD can be applied to a model-space in which the spin-orbital partners, $j_>(= l+1/2)$ and $j_<(= l- 1/2)$, associated with the same orbital quantum number $l$ is present. The $^{48}$Ca belongs to $fp-$model-space, which has spin-orbit partners. Thus, the present study of $0\nu\beta\beta$ of $^{48}$Ca using STD is of great interest. 

In the present work, we examine the roles of C, SO, and T components of two nucleon interaction GXPF1A~\cite{PhysRevC.69.034335, Honma2005} of $pf$-shell in the NMEs $2\nu\beta\beta$ and $0\nu\beta\beta$ of $^{48}$Ca. The NMEs for  $2\nu\beta\beta$ are calculated nonclosure approach. For  $0\nu\beta\beta$, NMEs are calculated in both closure approximation, and nonclosure approach using four different methods, namely, closure, running closure, running nonclosure, and the mixed method.

This paper is organized as follows. In sec. \ref{sec:II}, the expression of NMEs for $2\nu\beta\beta$ and the light neutrino-exchange mechanism of $0\nu\beta\beta$ are given. Discussion of closure approximation and nonclosure approach in different methods of NMEs calculations are given in sec. \ref{sec:III}. The details of the spin tensor-decomposition are given in sec. \ref{sec:IV}. The results and discussion are presented in sec. \ref{sec:V}. This work is summarized in sec. \ref{sec:VI}.
\section{\label{sec:II}Nuclear Matrix Elements}
\subsection{$2\nu\beta\beta$ mode}
The half-life of $2\nu\beta\beta$ of $0^+$ ground state (g.s) to $0^+$ g.s. transition is given by \cite{PhysRevC.75.034303,suhonen1998weak,faessler1998double,elliott2004double}
\begin{equation}
T_{1/2}^{2\nu,0}=F_{0}^{2\nu}|M_{GT}^{2\nu}(0^{+})|^{2},
\end{equation}
where $F_{0}^{2\nu}$ is the phase-space factor \cite{suhonen1998weak}. Here, only Gamow-Teller type NMEs ($M_{GT}^{2\nu}(0^{+})$) is relevant which can be written as \cite{PhysRevC.75.034303}
\begin{equation}
    M_{GT}^{2\nu}(0^+)=\sum_{k}\frac{\langle f||\sigma\tau^-||k\rangle\langle k||\sigma\tau^-||i\rangle}{E_k^{*}+E_0},
\end{equation}
where $\tau^-$ is the isospin lowering operator, in the present work $|i\rangle$ is $0^+$ ground state (g.s) of parent nucleus $^{48}$Ca, $|f\rangle$ is $0^+$ g.s. state of grand-daughter nucleus $^{48}$Ti, $|k\rangle$ is $1^+$ states of intermediate nucleus $^{48}$Sc, $E_k^{*}$ is the excitation energy of the $1^{+}$ states of $^{48}$Sc, and the constant $E_0$ is given by
\begin{equation}
    E_0=\frac{1}{2}Q_{\beta \beta}(0^+)+\triangle M.
\end{equation}
Here $Q_{\beta \beta}(0^+)$ is the Q value corresponding to $\beta \beta$ decay of $^{48}$Ca and $\triangle M$ is mass difference of $^{48}$Sc and $^{48}$Ca isotopes.

The reduced transition matrix elements ($\langle k||\sigma\tau^{-}||i\rangle$) can be written as a sum over product of one body transition densities (OBTD) and one body matrix elements ($\langle k'_{1}||\sigma\tau^{-}||k_{1}\rangle$)  \cite{brown2005lecture}
\begin{eqnarray}
\langle k||\sigma\tau^{-}||i\rangle=\sum_{k_{1}k'_{1}}\text{OBTD}(k,i,k'_{1},k_{1},J_{k})\times\langle k'_{1}||\sigma\tau^{-}||k_{1}\rangle\nonumber,
\\
\end{eqnarray}
where $k_1$ refers to the set of quantum numbers $(n_1, l_1,j_1)$, and $J_k$ is the spin of intermediate state $|k\rangle$.
OBTD in proton-neutron formalism can be written as \cite{PhysRevC.88.064312}
\begin{equation}
\label{eq:obtd}
  \text{OBTD}(k,i,k'_{1},k_{1},\mathcal{J})=\frac{\langle k||[a_{k'_{1}}^{+}\otimes\widetilde{a}_{k_{1}}]_\mathcal{J}||i\rangle}{\sqrt{2\mathcal{J}+1}}, 
\end{equation}
where $a_{k'_{1}}^{+}$ and $\widetilde{a}_{k_{1}}$ are the one particle creation and annihilation operator, respectively. 

One body matrix elements (OBMEs) are given by \cite{brown2005lecture}
\begin{eqnarray}
&&\langle k'_{1}||\sigma\tau^{-}||k_{1}\rangle\nonumber\\
=&&(-1)^{l'_{1}+j'_{1}+3/2}\sqrt{(2j'_{1}+1)(2j_{1}+1)}\left\{ \begin{array}{ccc}
1/2 & 1/2 & 1\\
j_{1} & j'_{1} & l'_{1}
\end{array}\right\}\nonumber\\
&&\times\langle s||\vec{s}||s\rangle\delta_{l'_{1},l_{1}}\delta_{n'_{1},n_{1}}
\end{eqnarray}
\subsection{$0\nu\beta\beta$ mode}
The decay rate for light neutrino-exchange mechanism of $0\nu\beta\beta$  can be written as \cite{RevModPhys.80.481,vergados2012theory}
\begin{equation}
    [T^{0\nu}_\frac{1}{2}]^{-1}=G^{0\nu}|M^{0\nu}|^2(\frac{m_{\beta\beta}}{m_e})^2,
\end{equation}
where $G^{0\nu}$ is well-known phase-space factor \cite{PhysRevC.85.034316}, $M^{0\nu}$ is the nuclear
matrix element, and $m_{\beta\beta}$ is the effective Majorna neutrino mass defined by the neutrino mass eigenvalues $m_k$ and the neutrino mixing matrix elements $U_{ek}$:
\begin{equation}
    \langle m_{\beta\beta}\rangle=\lvert \sum_km_kU_{ek}^2\rvert.
\end{equation}
The nuclear matrix element $M^{0\nu}$ can be expressed as the sum of Gamow-Teller ($M^{0\nu}_{GT}$), Fermi($M^{0\nu}_{F}$), and tensor ($M^{0\nu}_{T}$) matrix elements \cite{RevModPhys.80.481}
\begin{equation}
    M^{0\nu}=M^{0\nu}_{GT}-\left(\frac{g_V}{g_A}\right)^{2}M^{0\nu}_{F}+M^{0\nu}_{T},
\end{equation}
where $g_V$ and $g_A$ are the vector and axial-vector constant, respectively. In our calculation $g_V=1$ and bare unquenched $g_A=1.27$ was used \cite{PhysRevC.101.014307}. $M^{0\nu}_{GT}$, $M^{0\nu}_{F}$ and $M^{0\nu}_{T}$ matrix elements of the scalar two-body transition operator $\mathcal{O}_{12}^\alpha$  of $0\nu\beta\beta$
can be expressed as \cite{PhysRevLett.113.262501}:
\begin{eqnarray}
\label{Eq:NMEMAIN}
&&M^{0\nu}_{\alpha}=\langle f|\tau_{-1}\tau_{-2}\mathcal{O}_{12}^\alpha|i\rangle,
\end{eqnarray}
where $\alpha={(F,GT,T)}$, in the present work $|i\rangle$ is $0^+$ ground state (g.s) of parent nucleus $^{48}$Ca, $|f\rangle$ is $0^+$ g.s. state of grand-daughter nucleus $^{48}$Ti, $\tau_{-}$ is the isospin annihilation operator. Scalar two-particle transition operators of $0\nu\beta\beta$ containing spin and  radial neutrino potential operators are given by \cite{PhysRevC.88.064312}:
\begin{eqnarray}
\mathcal{O}_{12}^{GT}&&=\tau_{1-}\tau_{2-}(\mathbf{\sigma_1.\sigma_2)}H_{GT}(r,E_k),
\nonumber\\
\label{eq:opnc}
\mathcal{O}_{12}^{F}&&=\tau_{1-}\tau_{2-}H_{F}(r,E_k),
\\
\mathcal{O}_{12}^{T}&&=\tau_{1-}\tau_{2-}S_{12}H_{T}(r,E_k),
\nonumber
\end{eqnarray}
where $S_{12}=3(\mathbf{\sigma_1 .\hat{r})(\sigma_2.\hat{r})-(\sigma_1.\sigma_2)}$, $\mathbf{r=r_1-r_2}$, and $r=|\mathbf{r}|$ is inter nucleon distance of the decaying nucleons.

The neutrino potential for light-neutrino exchange mechanism of $0\nu\beta\beta$ are given as integral over Majorana neutrino momentum q \cite{PhysRevC.88.064312}:
\begin{equation}
\label{eq:npnc}
H_\alpha (r,E_{k})=\frac{2R}{\pi}\int_{0}^{\infty}\frac{j_p(qr)
h_\alpha(q^2)qdq}{q+E_{k}-(E_{i}+E_{f})/2},
\end{equation}
where $j_p(qr)$ is spherical Bessel function, variable p=0 for $M_{GT}^{0\nu}$ and $M_{F}^{0\nu}$, and p=2 for $M_{T}^{0\nu}$, R is the radius of the parent nucleus, $E_k$ is the energy of intermediate states, $E_i$ is the energy of the initial state, $E_f$ is the energy of the final state, and $h_\alpha(q^2)$ is the form factors that incorporates the effects of higher-order currents (HOC) and finite nucleon size (FNS) \cite{PhysRevC.60.055502}. Form factors $h_\alpha(q^2)$ used in our calculations are given in Ref. \cite{PhysRevC.101.014307}.

The short range nature of the two-nucleon interaction is taken care by multiplying relative harmonic oscillator wavefunciton $\psi_{nl}$ in radial integral $\langle n',l'|H_\alpha(r)|n,l\rangle$ with a correlation function $f(r)$ \cite{PhysRevC.81.024321};
\begin{equation}
    \psi_{nl}(r)\longrightarrow[1+f(r)]\psi_{nl}(r),
\end{equation}
where $f(r)$ can be parametrized as \cite{PhysRevC.79.055501}
\begin{equation}
    f(r)=-ce^{ar^2}(1-br^2).
\end{equation}
The parameters $a$, $b$ and $c$ for Miller-Spencer, CD-Bonn and AV18 type SRC parametrization are given in Refs. \cite{PhysRevC.81.024321,PhysRevC.101.014307}.
\section{\label{sec:III}Closure, nonclosure, and mixed methods}
If one replaces the energies of intermediate states in Eq.~(\ref{eq:npnc}) by an average constant value, one reaches the closure approximation:
\[[E_{k}-(E_{i}+E_{f})/2]\rightarrow \langle E\rangle.\]
In closure approximation, neutrino potential of Eq.~(\ref{eq:npnc}) becomes \cite{PhysRevC.81.024321}
\begin{equation}
\label{eq:npc}
H_\alpha (r)=\frac{2R}{\pi}\int_{0}^{\infty}\frac{j_p(qr)
h_\alpha(q^2)qdq}{q+\langle E \rangle,}
\end{equation}
and the transition operators of $0\nu\beta\beta$ of Eq.~(\ref{eq:opnc}) can be re-written as
\begin{eqnarray}
O_{12}^{GT}&&=\tau_{1-}\tau_{2-}(\mathbf{\sigma_1.\sigma_2)}H_{GT}(r),
\nonumber\\
\label{eq:opc}
O_{12}^{F}&&=\tau_{1-}\tau_{2-}H_{F}(r),
\\
O_{12}^{T}&&=\tau_{1-}\tau_{2-}S_{12}H_{T}(r).
\nonumber
\end{eqnarray}
Closure approximation has significant advantage over nonclosure approach because it eliminates the complexity of calculating large number of intermediate states which can be computationally challenging for heavy nuclear systems. This approximation is also very good as the values of q that dominate the matrix elements are of the order of 100–200 MeV, while the relevant excitation energies of the intermediate states are only of the order of 10 MeV \cite{PhysRevC.88.064312}. The most important part of closure approximation is to use a suitable value of average closure energy $\langle E \rangle$ that will take care the combine effects of a large number of intermediate states. In the present work, we have used standard closure energy $\langle E\rangle=7.72$ MeV \cite{PhysRevC.81.024321,PhysRevC.88.064312}.

In nonclosure approach, one needs to calculate the neutrino potential of Eq.~(\ref{eq:npnc}) explicitly in terms of energy $E_k$ of large number of virtual intermediate states $|k\rangle$ (for our case ($^{48}$Sc). For our nonclosure calculations, we have used \cite{PhysRevC.88.064312}
\begin{equation}
    E_{k}-(E_{i}+E_{f})/2\rightarrow 1.9 \text{MeV}+E_k^{*},
\end{equation}
where $E_k^{*}$ is the excitation energy of the intermediate states $|k\rangle$.

In the present work, based on the closure approximation and nonclosure approach, we have used four different methods, namely, closure, running closure, running nonclosure, and mixed method to calculate NMEs of $0\nu\beta\beta$. Descriptions of the above methods are given below.

\textbf{Closure method:}
In closure method, the NMEs are calculated using $0\nu\beta\beta$ transition operator of Eq.~(\ref{eq:opc}) and the neutrino potential of the closure approximation defined in Eq.~(\ref{eq:npc}). In this method NMEs defined in~(\ref{Eq:NMEMAIN}) can be written as sum over products of two-nucleon transfer amplitudes (TNAs) and anti-symmetric two-body matrix elements ($\langle k_1',k_2',JT|\tau_{-1}\tau_{-2}{O}_{12}^\alpha|k_1,k_2,JT\rangle_A$):
\begin{eqnarray}
\label{Eq:NMETNA}
&&\mathcal{M}^{0\nu}_{\alpha}=\sum_{m,J,k_1'\leqslant k_2',k_1\leqslant k_2}\text{TNA}(f,m,k_1', k_2', J_m)\nonumber\\
&&\text{TNA}(i,m,k_1, k_2, J_m)\times\langle k_1',k_2':JT|\tau_{-1}\tau_{-2}{O}_{12}^\alpha|k_1,k_2:JT\rangle_A,\nonumber\\
\end{eqnarray}
where $k$ stands for the set of spherical quantum numbers $(n; l; j)$, and A denotes that the two-body matrix elements are obtained using anti-symmetric two-nucleon wavefunction. In our study, $|i\rangle$ is $0^+$ g.s. of the parent nucleus $^{48}$Ca, $|m\rangle$ is the large number of states of intermediate nucleus ($^{46}$Ca) with allowed spin-parity ($J^{\pi}$),  $|f\rangle$ is the $0^+$ g.s. of the granddaughter nucleus $^{48}$Ti, and $k$ has the spherical quantum numbers for $0f_{7/2}$, $0f_{5/2}$, $1p_{3/2}$, and $1p_{1/2}$ orbitals. Complete expression of anti-symmetric two-body matrix elements (TBMEs) is given in Refs. \cite{PhysRevC.101.014307,PhysRevC.81.024321}.

TNA is calculated with a large set of intermediate states $|m\rangle$ of the (n-2) nucleons system ($^{46}$Ca in the present study), where n is the number of nucleons for the parent nucleus.
TNA is given by \cite{PhysRevLett.113.262501}
\begin{equation}
  \text{TNA}(f,m,k_1', k_2', J_m)=\frac{\langle f||A^+(k_1', k_2' ,J_m)||m\rangle}{\sqrt{2J_0+1}}.
\end{equation}
Here, 
\begin{equation}
  A^+(k_1', k_2' ,J)=\frac{[ a^{+}(k_1')\otimes a^{+} (k_2')]^{J}_{M}}{\sqrt{1+\delta_{{k_1'}{k_2'}}}}
\end{equation}
is the two particle creation operator of rank $J$, $J_m$ is the spin of the allowed states of $^{46}$Ca, $J_0$ is spin of $|i\rangle$ and $|f\rangle$. In Eq. (\ref{Eq:NMETNA}), $J_m$=$J$ when $J_0$=0 \cite{PhysRevLett.113.262501}. 
The TNA is normalized such that $\text{TNA}^2=n_p(n_p-1)/2$ for the removal of two protons and $\text{TNA}^2=n_n(n_n-1)/2$ for the removal of two neutrons, where $n_{p(n)}$ are the total number of protons (neutrons) in the model-space \cite{PhysRevLett.113.262501}.

\textbf{Running closure method:}
In running closure method, one uses the same $0\nu\beta\beta$ transition operator and neutrino potential as closure method. However, in this method one gets the true virtual intermediate nucleus after one neutron from parent nucleus decay into one proton. In the present study $^{48}$Sc is the true virtual intermediate nucleus. The NMEs for running closure method in proton-neutron formalism can be written as sum over product of one body transition densities (OBTD) and  reduced non anti-symmetric two body matrix elements ($\langle k_1',k_2':J||\tau_{-1}\tau_{-2}{O}_{12}^\alpha||k_1,k_2:J\rangle$) \cite{PhysRevC.88.064312}:
\begin{eqnarray}
\label{eq:nmerc}
&&\mathcal{M}_{\alpha}^{0\nu}(E_c)=\sum_{k'_{1}k'_{2}k_{1}k_{2}JJ_{k}}\sum_{E^{*}_{k}\leqslant E_c}\sqrt{(2J_{k}+1)(2J_{k}+1)(2J+1)}\nonumber\\
&&\times(-1)^{j_{k1}+j_{k2}+J}
\left\{ \begin{array}{ccc}
j_{k1^{'}} & j_{k1} & J_{k}\\
j_{k2} & j_{k2^{'}} & J
\end{array}\right\}\text{OBTD}(k,f,k'_{2},k_{2},J_{k})\nonumber\\ &&\times \text{OBTD}(k,i,k'_{1},k_{1},J_{k})\langle k_1',k_2':J||\tau_{-1}\tau_{-2}{O}_{12}^\alpha||k_1,k_2:J\rangle.\nonumber\\
\end{eqnarray}
Here $k_1$ represents set of spherical quantum numbers ($n_1,l_1,j_1$) for a orbital, $J$ is the coupled spin of two decaying neutrons or two final created protons, $J_k$ is allowed spin of the intermediate nucleus, $E^{*}_k$ is the excitation energy of each allowed $J_{k}^{\pi}$ of intermediate nucleus which can run upto cutoff excitation energy $E_{c}$ in the summation. Considering states whose excitation energy $E^{*}_k$ goes up to $E_{c}$ gives almost constant NMEs when $E_{c}$ is large enough. The OBTD are calculated using Eq.~(\ref{eq:obtd}) with large number of virtual intermediate states of $^{48}$Sc for all allowed spin-parity.

\textbf{Running nonclosure method:}
In running nonclosure method, NMEs are calculated with the nonclosure $0\nu\beta\beta$ transition operators given in Eq.~(\ref{eq:opnc}). Nonclosure neutrino potential defined in Eq.~(\ref{eq:npnc}) are calculated explicitly in terms of excitation energy of large number allowed states of intermediate nucleus ($^{48}$Sc). The NMEs for running nonclosure in proton-neutron formalism can be defined as \cite{PhysRevC.88.064312} 
\begin{eqnarray}
\label{eq:nmernc}
&&M_{\alpha}^{0\nu}(E_c)=\sum_{k'_{1}k'_{2}k_{1}k_{2}JJ_{k}}\sum_{E^{*}_{k}\leqslant E_c}\sqrt{(2J_{k}+1)(2J_{k}+1)(2J+1)}\nonumber\\
&&\times(-1)^{j_{k1}+j_{k2}+J}
\left\{ \begin{array}{ccc}
j_{k1^{'}} & j_{k1} & J_{k}\\
j_{k2} & j_{k2^{'}} & J
\end{array}\right\}\text{OBTD}(k,f,k'_{2},k_{2},J_{k})\nonumber\\ &&\times \text{OBTD}(k,i,k'_{1},k_{1},J_{k})\langle k_1',k_2':J||\tau_{-1}\tau_{-2}{\mathcal{O}_{12}^\alpha}||k_1,k_2:J\rangle\nonumber\\
\end{eqnarray}
Complete expression of non anti-symmetric reduced TBMEs for running nonclosure and running closure method is given in Ref. \cite{PhysRevC.88.064312}.

\textbf{Mixed method:}
The mixed method is the superposition of running nonclosure, running closure, and closure method. NMEs in the mixed method is written as \cite{PhysRevC.88.064312}
\begin{equation}
   \bar{M}_{\alpha}^{0\nu}(E_c)=M_{\alpha}^{0\nu}(E_c)-\mathcal{M}_{\alpha}^{0\nu}(E_c)+\mathcal{M}^{0\nu}_{\alpha}
\end{equation}
Mixed methods has very good convergence property \cite{PhysRevC.88.064312}. Thus, this method is  particularly useful for calculating NMEs for higher mass region isotopes. Because of high convergence, NMEs calculated with few states of intermediate nucleus can give almost constant NMEs.
\section{\label{sec:IV}Spin-Tensor Decomposition}
In the present study, we have employed spin-tensor decomposition \cite{dirim1968jp,Kirson:1973ffz,RevModPhys.48.191,PhysRevC.15.1483,kenji1980spin,smirnova2010shell,brown1985spin,PhysRevC.45.662,PhysRevC.86.034314} to decompose GXPF1A interaction into its central (C), spin-orbit (SO), and tensor (T) force components. In spin-tensor decomposition, the interaction between two-nucleon is defined as the linear sum of the scalar product of configuration space operator $Q$ and spin space operator $S$ of rank $k$ \cite{Kirson:1973ffz}:
\begin{equation}
\label{3}
V = \sum_{k = 0}^{2} V(k) = \sum_{k = 0}^{2} Q^{k}.S^{k},
\end{equation} where rank $k$ = 0, 1  and 2 represent central, spin-orbit and tensor force, respectively. Using the $LS$-coupled two-nucleon wave functions, the matrix element for each $V(k)$ can be calculated from the matrix element of $V$ \cite{dirim1968jp}:
\begin{equation}
\label{4}
\begin{split}
\langle (ab),LS;J|V(k)|(cd),L'S';J \rangle = (2k+1) (-1)^{J} \\
\times
\left \{
  \begin{tabular}{ccc}
  L & S & J \\
  S$'$ & L$'$ & k \\
  \end{tabular}
\right \}
\sum_{J'} (-1)^{J'} (2J'+1)
\left \{
  \begin{tabular}{ccc}
  L & S & J$'$ \\
  S$'$ & L$'$ & k \\
  \end{tabular}
\right \}
\\ \times\langle (ab),LS;J'|V|(cd),L'S';J' \rangle,
\end{split}
\end{equation}where $a$ refers to the set of quantum numbers ($n_{a}$, $l_{a}$).
\section{\label{sec:V}Results and Discussion}
The required TNA and OBTD are calculated using shell-model code NushellX@MSU \cite{brown2014shell}. To calculate TNA, we have considered the first 100 states of intermediate nucleus $^{46}$Ca for each allowed spin-parity ($J^\pi$). Considering first 100 states gives alomst constant NMEs \cite{PhysRevC.101.014307,PhysRevLett.113.262501}

\begin{table}[h]
\caption{\label{tab:tabledbd} NMEs for $2\nu\beta\beta$ of $^{48}$Ca, calculated with different components (C, C+SO, and C+SO+T) of GXPF1A interaction.}
\begin{ruledtabular}
\begin{tabular}{cccc}
NME&C&C+SO&C+SO+T\\ \hline
$M^{2\nu}$
&0.070  &-0.065  &0.059
\\
\end{tabular}
\end{ruledtabular}
\end{table}
\begin{figure}[t]
\centering
\includegraphics[trim=2cm 1cm 2cm 2cm,width=\linewidth]{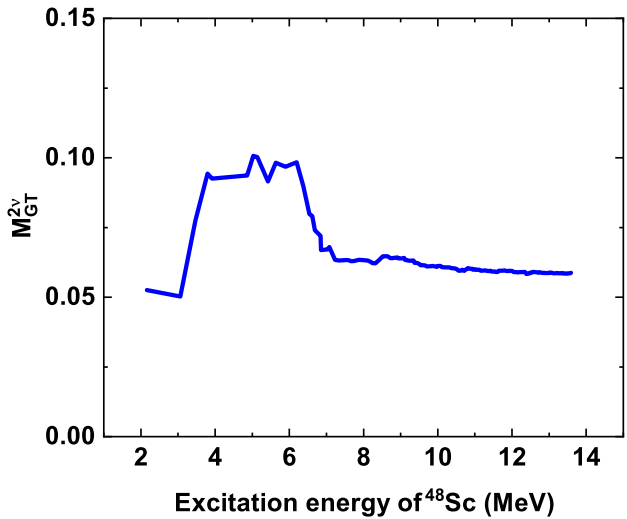}
\caption{\label{fig:nmevsexcdbd}(Color online) Variation of NMEs for $2\nu\beta\beta$ of $^{48}$Ca with excitation energy of $1^{+}$ states of the virtual intermediate nucleus $^{48}$Sc.}
\end{figure}
\begin{figure}[b]
\centering
\includegraphics[trim=2cm 1cm 2cm 2cm,width=\linewidth]{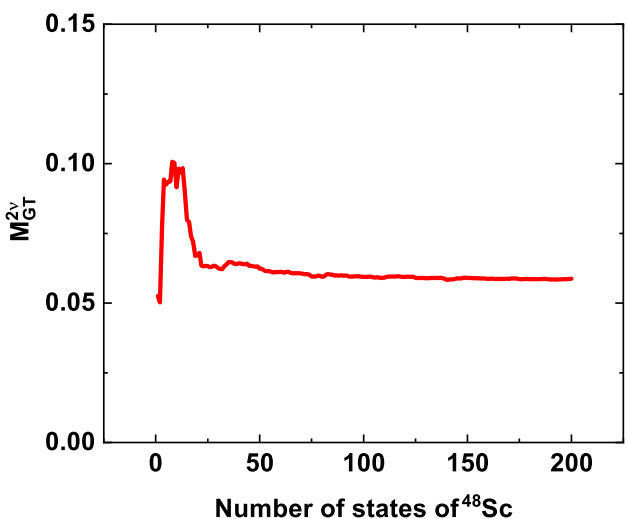}
\caption{\label{fig:nmevsnumdbd}(Color online) Variation of NMEs for $2\nu\beta\beta$ of $^{48}$Ca with number of $1^{+}$ states of the virtual intermediate nucleus $^{48}$Sc.}
\end{figure}
The OBTD are calculated by considering first 150 states for each allowed spin-parity ($J_{k}^{\pi}$) of virtual intermediate nucleus ($^{48}$Sc in our case) for both $2\nu\beta\beta$, and $0\nu\beta\beta$ of $^{48}$Ca. Earlier it was found that considering first 100 states for each $J_k$ of $^{48}$Sc to calculate OBTD gives NMEs with uncertainty about 1$\%$ \cite{PhysRevC.88.064312}. The calculation of the one body matrix elements for $2\nu\beta\beta$ and two-body matrix elements (TBMEs) of $0\nu\beta\beta$ was done using the program written by us. 
\begin{table*}
\caption{\label{tab:tablendbd} NMEs for $0\nu\beta\beta$ (light neutrino-exchange mechanism) of $^{48}$Ca, calculated in closure, running closure, running nonclosure, and mixed methods with different components (C, C+SO and C+SO+T) of GXPF1A interaction for different SRC parametrization. Closure energy $\langle E\rangle =$7.72 MeV was used for closure and running closure methods.}
\begin{ruledtabular}
\begin{tabular}{cccccccccccccc}
 &&\multicolumn{3}{c}{Closure}&\multicolumn{3}{c}{Running closure}&\multicolumn{3}{c}{Running nonclosure}&\multicolumn{3}{c}{Mixed}
  \vspace{.1cm}\\
  \cline{3-5}
  \cline{6-8}
  \cline{9-11}
   \cline{12-14}
   \vspace{.1cm}
NME&SRC&C&C+SO&C+SO+T&C&C+SO&C+SO+T&C&C+SO&C+SO+T&C&C+SO&C+SO+T\\ \hline
$M_F^{0\nu}$&None
&-0.261&0.223&-0.207
&-0.258&0.220&-0.206
&-0.263&0.224&-0.210
&-0.266&0.227&-0.211

\\
$M_F^{0\nu}$&Miller-Spencer
&-0.185&0.152&-0.141
&-0.183&0.151&-0.141
&-0.186&0.153&-0.143
&-0.188&0.154&-0.143

\\
$M_F^{0\nu}$&CD-Bonn
&-0.279&0.239&-0.222
&-0.276&0.236&-0.221
&-0.282&0.241&-0.226
&-0.285&0.244&-0.227

\\
$M_F^{0\nu}$&AV18
&-0.258&0.219&-0.204
&-0.255&0.216&-0.203
&-0.261&0.221&-0.207
&-0.264&0.224&-0.208

\\
\\
$M_{GT}^{0\nu}$&None
&0.841&-0.739&0.709
&0.864&-0.785&0.707
&0.942&-0.853&0.778
&0.919&-0.807&0.780

\\
$M_{GT}^{0\nu}$&Miller-Spencer
&0.590&-0.506&0.491
&0.614&-0.552&0.489
&0.682&-0.612&0.551
&0.658&-0.566&0.553

\\
$M_{GT}^{0\nu}$&CD-Bonn
&0.873&-0.767&0.736
&0.897&-0.814&0.734
&0.978&-0.886&0.808
&0.954&-0.839&0.810

\\
$M_{GT}^{0\nu}$&AV18
&0.801&-0.701&0.673
&0.825&-0.748&0.672
&0.903&-0.817&0.743
&0.879&-0.770&0.744

\\
\\
$M_T^{0\nu}$&None
&-0.079&0.069&-0.075
&-0.076&0.067&-0.072
&-0.079&0.070&-0.074
&-0.082&0.072&-0.077

\\
$M_T^{0\nu}$&Miller-Spencer
&-0.080&0.070&-0.076
&-0.078&0.068&-0.073
&-0.080&0.071&-0.075
&-0.082&0.073&-0.078

\\
$M_T^{0\nu}$&CD-Bonn
&-0.081&0.071&-0.077
&-0.078&0.069&-0.074
&-0.081&0.072&-0.076
&-0.084&0.074&-0.079

\\
$M_T^{0\nu}$&AV18
&-0.081&0.071&-0.077
&-0.079&0.069&-0.074
&-0.081&0.072&-0.076
&-0.083&0.074&-0.079

\\
\\
$M^{0\nu}$&None
&0.924&-0.808&0.763
&0.948&-0.854&0.763
&1.026&-0.923&0.834
&1.002&-0.877&0.834

\\
$M^{0\nu}$&Miller-Spencer
&0.625&-0.530&0.502
&0.650&-0.577&0.503
&0.718&-0.636&0.565
&0.693&-0.589&0.564

\\
$M^{0\nu}$&CD-Bonn
&0.965&-0.844&0.797
&0.989&-0.891&0.797
&1.071&-0.964&0.872
&1.047&-0.917&0.872

\\
$M^{0\nu}$&AV18
&0.880&-0.765&0.723
&0.904&-0.812&0.723
&0.984&-0.882&0.796
&0.960&-0.835&0.796

\\
\end{tabular}
\end{ruledtabular}
\end{table*}
\begin{figure*}
\centering
\includegraphics[trim=3cm 8.5cm 2.5cm 2cm,height=8.5cm,width=\linewidth]{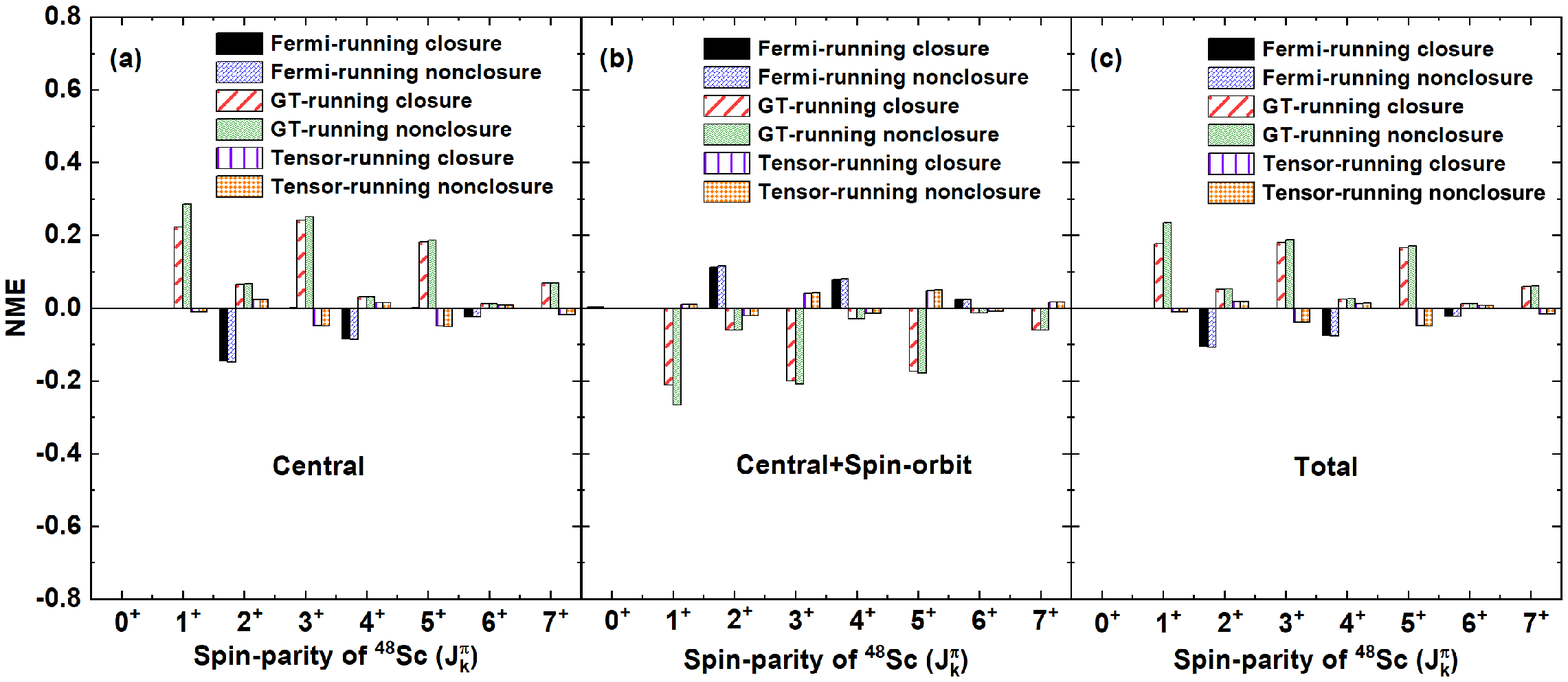}
\caption{\label{fig:NMEvsJk}(Color online) NMEs of $0\nu\beta\beta$ (light neutrino-exchange mechanism) of $^{48}$Ca for different spin-parity ($J_{k}^{\pi}$) of intermediate nucleus $^{48}$Sc. NMEs are calculated in running closure and running nonclosure methods with (a) C (b) C+SO, and (c) Total (C+SO+T) of GXPF1A interaction for AV18 SRC parametrization. $\langle E\rangle=7.72$ MeV was used for running closure method.}
\end{figure*}
\begin{figure*}
\centering
\includegraphics[trim=3cm 7cm 2.5cm 2cm,height=8.5cm,width=\linewidth]{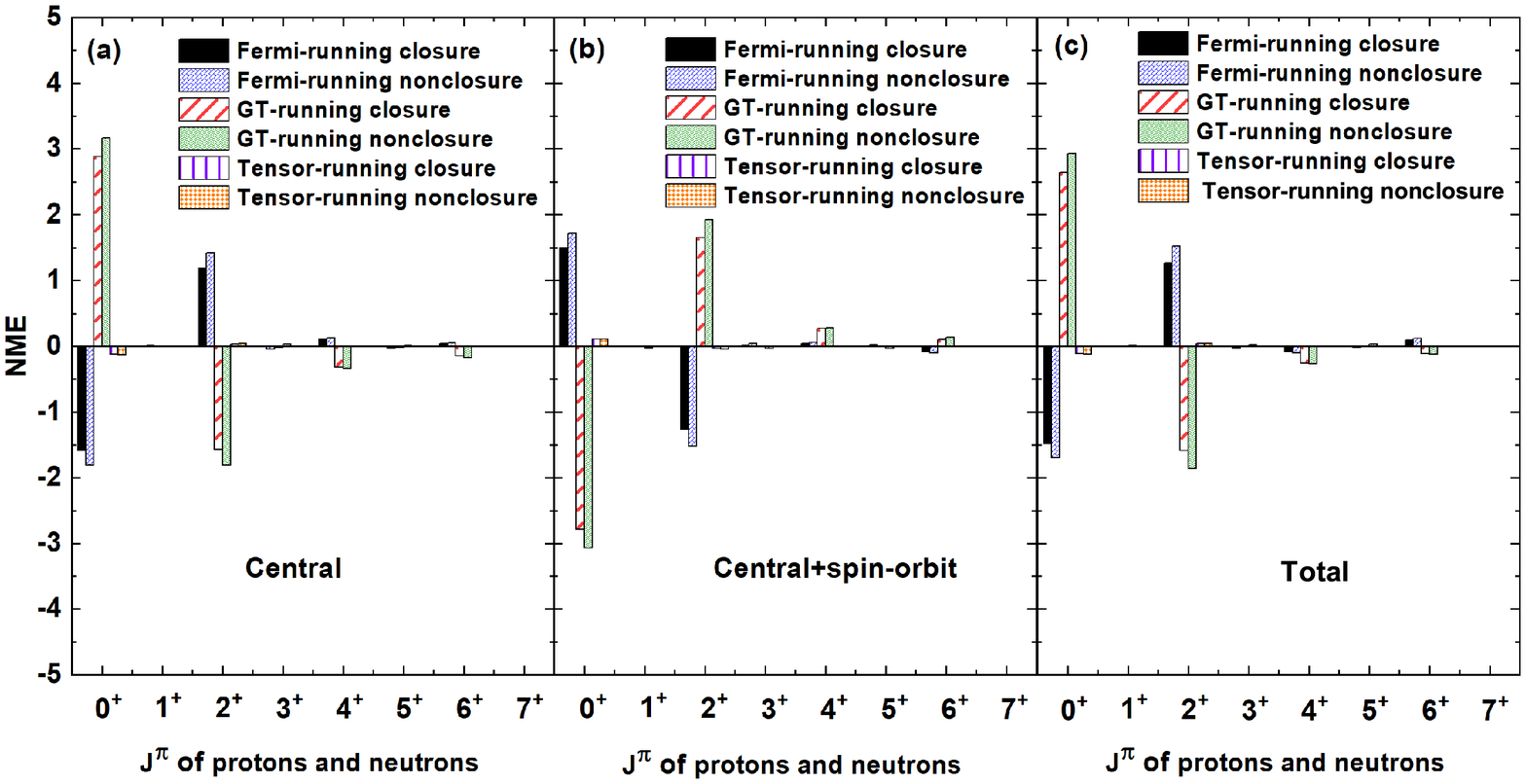}
\caption{\label{fig:NMEvsJ}(Color online) NMEs of $0\nu\beta\beta$ (light neutrino-exchange mechanism) of $^{48}$Ca for different coupled spin-parity ($J^{\pi}$) of two initial neutrons or two final created protons. NMEs are calculated in running closure and running nonclosure methods with (a) C (b) C+SO, and (c) Total (C+SO+T) of GXPF1A interaction for AV18 SRC parametrization. $\langle E\rangle=7.72$ MeV was used for running closure method.}
\end{figure*}

To examine the roles of C, SO and T components of two nucleon interaction GXPF1A, NMEs for both $2\nu\beta\beta$ and $0\nu\beta\beta$ are calculated first with C component of the interaction, then by adding SO component to C component and finally by adding T component to C+SO component which is same as the total interaction. Calculated NMEs of $2\nu\beta\beta$ with C, C+SO, and total (C+SO+T) of the GXPF1A interaction are given in Table ~\ref{tab:tabledbd}. 

From Table ~\ref{tab:tabledbd}, it is found that total NMEs of $2\nu\beta\beta$ calculated with C component of the interaction has positive magnitude. NMEs calculated by adding SO component to C component of the interaction is decreased by about 7$\%$ in magnitude and sign of NMEs is changed. By adding T component to C+SO component of the interaction, NMEs further decrease by about 9$\%$ in magnitude, and sign change of NMEs is again seen.

To study the dependence of NMEs for $2\nu\beta\beta$ on excitation energy of $1^+$ states of virtual intermediate nucleus $^{48}$Sc, we have shown the variation of NMEs with excitation energy in Fig. \ref{fig:nmevsexcdbd}. Here NMEs are calculated with total GXPF1A interaction. It is found that the first few low lying states up to around 10 MeV contribute constructively and destructively, and NMEs are mostly saturated and becomes constant at high excitation energy. In our calculation, we have considered states which go up to around 12 MeV, which gives an almost constant value of NMEs. A similar trend is seen with NMEs calculated with C and C+SO component of the interaction.

Variation of NMEs of $2\nu\beta\beta$ with the number of states of virtual intermediate nucleus $^{48}$Sc is given in Fig. \ref{fig:nmevsnumdbd}. Here NMEs are calculated with total GXPF1A interaction. It is observed that the first 50 low lying states contribute constructively and destructively, and at a large number of states, NMEs becomes constant. In our calculation, we have considered the first 150 $1^+$ states of $^{48}$Sc, which gives almost constant NMEs. A similar dependence of NMEs on the number of intermediate states is found with NMEs calculated by C, and C+SO component of the interaction.  

Calculated NMEs of $0\nu\beta\beta$ with C, C+SO components and total (C+SO+T) of GXPF1A interaction are given in Table ~\ref{tab:tablendbd}. Here NMEs are calculated in both closure approximation and nonclosure approach with four different methods: closure, running closure, running nonclosure, and mixed methods. NMEs are calculated, including the standard effects of FNS and HOC with different SRC parametrization.

It is found that total NMEs of $0\nu\beta\beta$ calculated with C component of the interaction for different SRC parametrization in all the methods have a positive value. Total NMEs calculated by adding SO component to C component of the interaction is decreased by about 13-15$\%$, 10-11$\%$, 10-11$\%$, 12-15$\%$, respectively, with closure, running closure, running nonclosure, and mixed methods for different SRC parametrization. The phase of the NMEs is also changed. By adding T component to the C+SO component of the interaction, the magnitude of the total NMEs is further decreased by about  5-6$\%$, 10-12$\%$, 9-11$\%$, 4-5$\%$, respectively, with closure, running closure, running nonclosure and mixed methods for different SRC parametrization. Phase changes are again seen. Similar trends of phase shift are also seen for Fermi, Gamow-Teller, and Tensor NMEs with all the methods. 

From Table~\ref{tab:tablendbd}, it is also found that for different SRC parametrization, the total NMEs calculated by C component of GXPF1A interaction with running closure method is about 2-4$\%$ larger than with closure method. In this case total NMEs in running nonclosure method is about 8-10$\%$ larger as compared to NMEs in running closure method and total NMEs in mixed method is decreased by about 2-3$\%$ as compared to running nonclosure method. 
\begin{figure*}[t]
\centering
\includegraphics[trim=2.5cm 2cm 3.5cm 2cm,width=\linewidth]{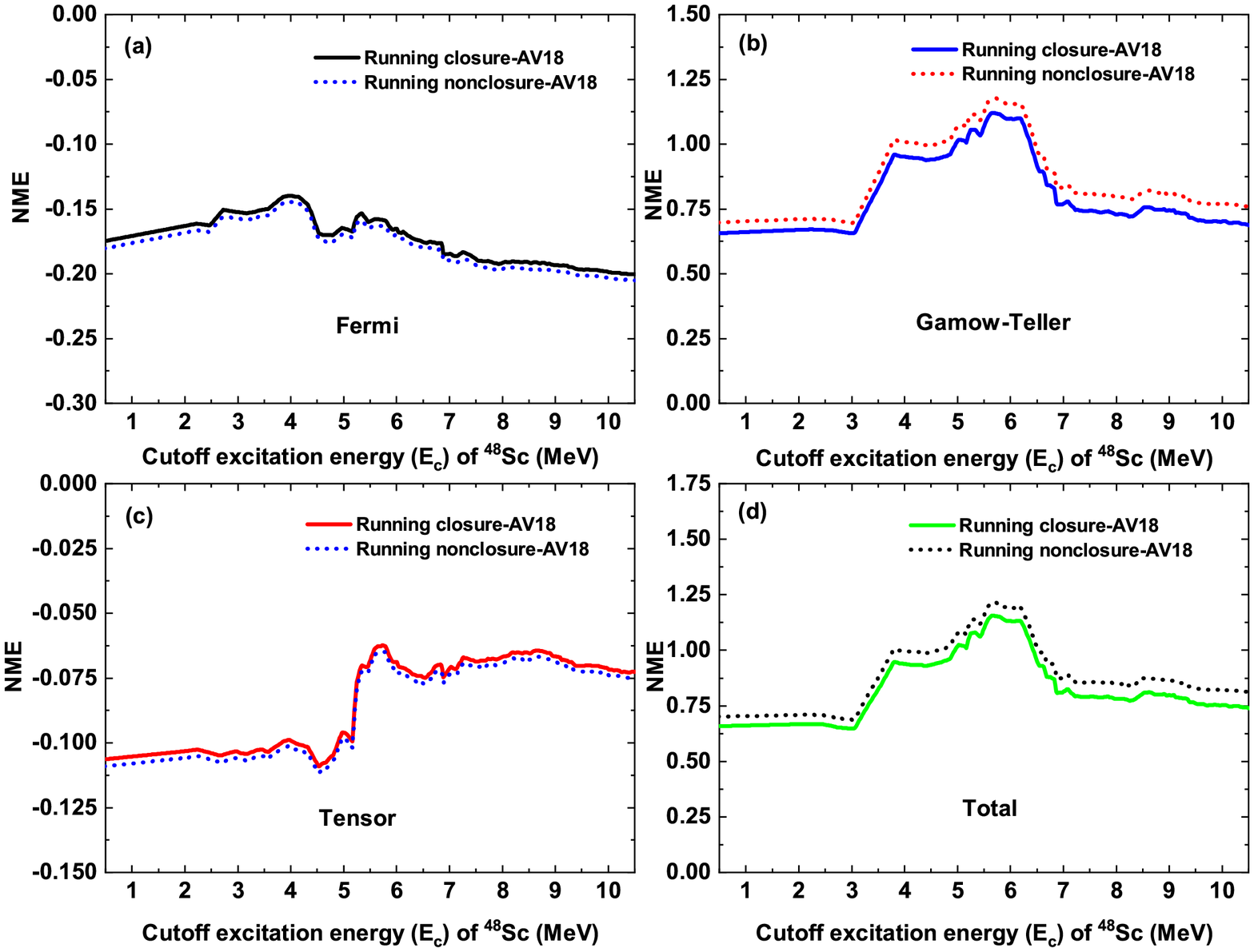}
\caption{\label{fig:nmevsek}(Color online) Variation of (a) Fermi (b) Gamow-Teller (c) tensor and (d) total NMEs for $0\nu\beta\beta$ (light neutrino-exchange mechanism) of $^{48}$Ca with cutoff excitation energy ($E_c$) of states of virtual intermediate nucleus $^{48}$Sc. NMEs are calculated with total GXPF1A interaction for AV18 SRC parametrization in running closure and running nonclosure methods. For running closure method, closure energy $\langle E\rangle $=7.72 MeV was used.}
\end{figure*}
\begin{figure*}[t]
\centering
\includegraphics[trim=2.5cm 2cm 3.5cm 2cm,width=\linewidth]{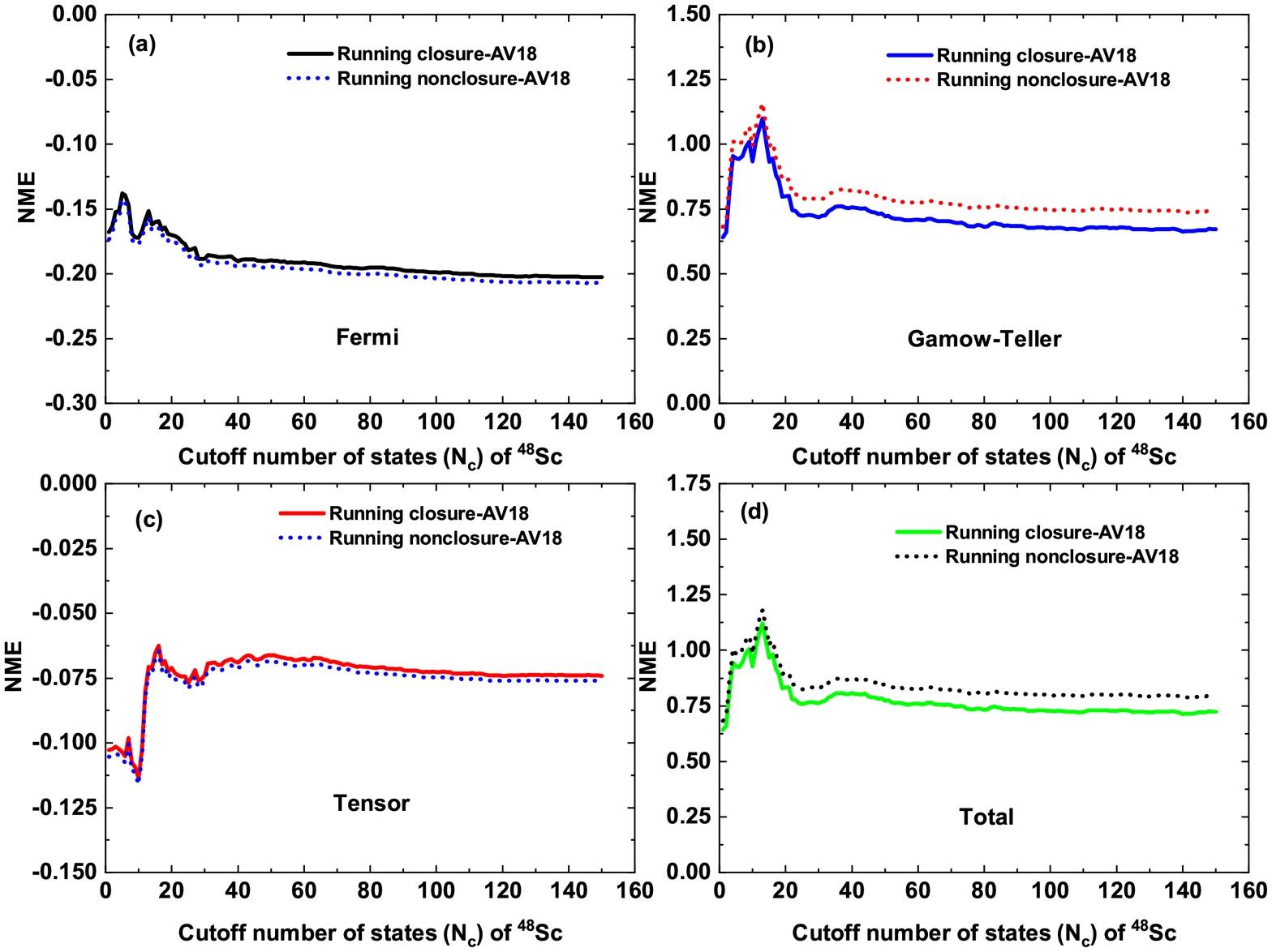}
\caption{\label{fig:nmevsnk}(Color online) Variation of (a) Fermi (b) Gamow-Teller (c) tensor and (d) total NMEs for $0\nu\beta\beta$ (light neutrino-exchange mechanism) of $^{48}$Ca with cutoff number of states ($N_c$) of virtual intermediate nucleus $^{48}$Sc. NMEs are calculated with total GXPF1A interaction for AV18 SRC parametrization in running closure and running nonclosure methods. For running closure method, closure energy $\langle E\rangle $=7.72 MeV was used.}
\end{figure*}

For different SRC parametrization, the magnitude of the total NMEs calculated by adding SO component to C component of the interaction in running closure method is about 6-9$\%$ larger than in the closure method. In this case, total NMEs in running nonclosure is about 8-10$\%$ larger as compared to NMEs in running closure method, and the total NMEs in the mixed method is about 5-7$\%$ smaller as compared running nonclosure method. 

Total NMEs calculated by adding T component to C+SO component of the interaction in running closure method is close to NMEs in the closure method. In this case, total NMEs in running nonclosure is about 9-12$\%$ more as compared to NMEs in closure and running closure method, and the total NMEs in the mixed method is close to NMEs in running nonclosure method for different SRC parametrization.

To study the dependence of NMEs in running closure and running nonclosure methods with spin-parity of the states of the intermediate nucleus $^{48}$Sc ($J_k^{\pi}$), coupled spin-parity of two initial protons or two final created protons ($J^\pi$), excitation energy of the states of $^{48}$Sc ($E_{k}^{*}$), and number of states of $^{48}$Sc ($N_k$), one can write the partial matrix elements in running closure method using Eq.(\ref{eq:nmerc}):
\begin{eqnarray}
\label{eq:rcpartial}
&&\mathcal{M}_{\alpha}^{0\nu}(J_{k},J,E_{k}^{*})=\sum_{k'_{1}k'_{2}k_{1}k_{2}}\sqrt{(2J_{k}+1)(2J_{k}+1)(2J+1)}\nonumber\\
&&\times(-1)^{j_{k1}+j_{k2}+J}
\left\{ \begin{array}{ccc}
j_{k1^{'}} & j_{k1} & J_{k}\\
j_{k2} & j_{k2^{'}} & J
\end{array}\right\}\text{OBTD}(k,f,k'_{2},k_{2},J_{k})\nonumber\\ &&\times \text{OBTD}(k,i,k'_{1},k_{1},J_{k})\langle k_1',k_2':J||\tau_{-1}\tau_{-2}{O_{12}^\alpha}||k_1,k_2:J\rangle,\nonumber\\
\label{eq:mjjkrc}
\end{eqnarray}
and the partial matrix elements in running nonclosure method using Eq. (\ref{eq:nmernc}) as
\begin{eqnarray}
\label{eq:rncpartial}
&&M_{\alpha}^{0\nu}(J_{k},J,E_{k}^{*})=\sum_{k'_{1}k'_{2}k_{1}k_{2}}\sqrt{(2J_{k}+1)(2J_{k}+1)(2J+1)}\nonumber\\
&&\times(-1)^{j_{k1}+j_{k2}+J}
\left\{ \begin{array}{ccc}
j_{k1^{'}} & j_{k1} & J_{k}\\
j_{k2} & j_{k2^{'}} & J
\end{array}\right\}\text{OBTD}(k,f,k'_{2},k_{2},J_{k})\nonumber\\ &&\times \text{OBTD}(k,i,k'_{1},k_{1},J_{k})\langle k_1',k_2':J||\tau_{-1}\tau_{-2}{\mathcal{O}_{12}^\alpha}||k_1,k_2:J\rangle\nonumber\\
\label{eq:mjjkrc}
\end{eqnarray}
Using Eq. (\ref{eq:rcpartial}) and (\ref{eq:rncpartial}), one can write NMEs as function of spin-parity ($J_{k}^{\pi}$) of the  states of intermediate nucleus $^{48}$Sc in running closure method as
\begin{eqnarray}
\mathcal{M}_{\alpha}^{0\nu}(E_c,J_k)=\sum_{J,E_{k}^{*}\leqslant E_c}\mathcal{M}_{\alpha}^{0\nu}(J_{k},J,E_{k}^{*})
\end{eqnarray}
 and in running nonclosure method as
\begin{eqnarray}
{M}_{\alpha}^{0\nu}(E_c,J_k)=\sum_{J,E_{k}^{*}\leqslant E_c}{M}_{\alpha}^{0\nu}(J_{k},J,E_{k}^{*})
\end{eqnarray}
Contributions of Fermi, Gamow-Teller, and tensor nuclear matrix elements through different $J_k^{\pi}$ of $^{48}$Sc are shown in Fig.~\ref{fig:NMEvsJk}. NMEs are calculated in running closure and running nonclosure methods with C, C+SO components and the total of GXPF1A interaction for AV18 SRC parametrization. It is found that for Fermi and Gamow-Teller type NMEs, contribution through each $J_{k}^{\pi}$ is coherent. But, contribution in tensor NMEs comes in opposite phase for different $J_{k}^{\pi}$ reducing the total tensor NMEs. 

Further, it is found that the dominating contribution comes from 2$^{+}$ state for Fermi type NMEs and with a small contribution from 4$^{+}$  and 6$^{+}$  states. Contributions through 0$^{+}$  and odd-$J_k^{\pi}$ states is negligible. All contributions of $J_k^{\pi}$ in Fermi type NMEs are negative with C component and total GXPF1A interaction, whereas contributions are positive for the C+SO component. Enhancement in Fermi type NMEs in running nonclosure method is mostly seen through 2$^{+}$ states as compared to running closure method.

For Gamow-Teller type NMEs, the dominating contribution comes through 1$^{+}$, 3$^{+}$, and 5$^{+}$ states. The small contribution comes through 2$^{+}$, 4$^{+}$, and 6$^{+}$ states and zero contribution from 0$^{+}$ states. All contributions of $J_k^{\pi}$ are positive for C component and total GXPF1A interactions, whereas all contributions are negative for the C+SO component of the interaction. Most dominating enhancements of running nonclosure NMEs are coming through 1$^{+}$ state for GT type NMEs as compared to running closure NMEs. 

For tensor type NMEs, coherent contribution comes through 1$^{+}$, 3$^{+}$, 5$^{+}$, and 7$^{+}$ states with contribution from 3$^{+}$, and 5$^{+}$ states being dominating. Contributions from 2$^{+}$, 4$^{+}$, and 6$^{+}$ states come with the opposite phase and reduce the total tensor NMEs. The phase of contributions of different $J_k^{\pi}$ is opposite for the C component and total GXPF1A interactions as compared with the C+SO component of the interaction. 

Similar pattern of variation of NMEs with different $J_{k}^{\pi}$ are seen with other type of SRC parametrization. 

We have also examined the variation of NMEs with coupled spin-parity ($J^\pi$) of two decaying neutrons and final created protons. One can write using Eq.~(\ref{eq:rcpartial}) and (\ref{eq:rncpartial}) NMEs as function of $J^\pi$ in running closure method as
\begin{eqnarray}
\mathcal{M}_{\alpha}^{0\nu}(E_c,J)=\sum_{J_k,E_{k}^{*}\leqslant E_c}\mathcal{M}_{\alpha}^{0\nu}(J_{k},J,E_{k}^{*})
\end{eqnarray}
and in running nonclosure method as
\begin{eqnarray}
{M}_{\alpha}^{0\nu}(E_c,J)=\sum_{J_k,E_{k}^{*}\leqslant E_c}{M}_{\alpha}^{0\nu}(J_{k},J,E_{k}^{*})
\end{eqnarray}
Contributions of NMEs through different $J^\pi$ is shown in Fig.~\ref{fig:NMEvsJ}. Here NMEs are calculated with C, C+SO and total of GXPF1A interaction in running closure and running nonclosure method for AV18 SRC parmaetrization. 

Unlike coherent contribution through different spin-parity ($J_k^\pi$) of the intermediate states in Fermi and Gamow-Teller NMEs, here contribution thorough different $J^\pi$ is not coherent for all type of NMEs. For all types of NMEs, the most dominating contribution comes from 0$^{+}$ states and 2$^{+}$ states. Also, the contribution from 0$^{+}$ and 2$^{+}$ states comes in opposite sign reducing the total NMEs. The small contribution comes through 4$^{+}$ and 6$^{+}$ states with almost negligible contributions from odd-$J^\pi$ states. Pairing effect is responsible for dominating even-$J^\pi$ contributions. It is found that contributions of different $J^\pi$ for C component and total GXPF1A interaction are of opposite sign as compared to NMEs with the C+SO component. Enhancement in NMEs with running nonclosure method are seen through $J^\pi$=0$^{+}$ and 2$^{+}$ states as compared to NMEs with running closure method. A similar dependence of NMEs with $J^\pi$ are seen for other SRC parametrizations. 

To study the dependence of NMEs with cutoff excitation energy ($E_c$) of the intermediate nucleus $^{48}$Sc, one can write the using Eq. (\ref{eq:rcpartial}), and (\ref{eq:rncpartial}) NMEs as function $E_c$ in running closure method as
\begin{eqnarray}
\mathcal{M}_{\alpha}^{0\nu}(E_c)=\sum_{J_k,J,E_{k}^{*}\leqslant E_c}\mathcal{M}_{\alpha}^{0\nu}(J_{k},J,E_{k}^{*})
\end{eqnarray}
and in running nonclosure method as
\begin{eqnarray}
{M}_{\alpha}^{0\nu}(E_c)=\sum_{J_k,J,E_{k}^{*}\leqslant E_c}{M}_{\alpha}^{0\nu}(J_{k},J,E_{k}^{*}).
\end{eqnarray}
Variation of Fermi, Gamow-Teller, tensor, and total NMEs with cutoff excitation energy ($E_c$) of $^{48}$Sc is shown in Fig. \ref{fig:nmevsek}. Here, NMEs are calculated in running closure and running nonclosure method with total GXPF1A interaction for AV18 type SRC parametrization. 
It is found that most of the contribution comes through the few low lying initial sates of $^{48}$Sc for each allowed spin-parity ($J_k^{\pi}$). At some high value of $E_c$, NMEs becomes almost constant. 
In Fig. \ref{fig:nmevsek}, we have shown the dependence of NMEs for $E_c$=0 to 10.5 MeV. Considering states whose excitation energy ($E_k^{*}$) goes up to 10.5 MeV gives NMEs with less than 1$\%$ uncertainty. NMEs are less sensitive with excitation energy of $^{48}$Sc because of the large neutrino momentum q, which is about $\sim$ 100-200 MeV sitting in the denominator of the neutrino potential in Eq. (\ref{eq:npnc}). A similar variation of NMEs with $E_c$ are found for other SRC parametrization and in NMEs calculated with C, and C+SO component of the GXPF1A interaction for different SRC parametrization. 

Instead of setting up cutoff on excitation energy ($E_c$) of $^{48}$Sc, one can also set a cutoff on the number of states ($N_c$) for each allowed $J_{k}^{\pi}$ of $^{48}$Sc to calculate the NMEs. 
One can write the NMEs as function of cutoff number of states ($N_c$) of $^{48}$Sc in running closure method as
\begin{eqnarray}
\mathcal{M}_{\alpha}^{0\nu}(N_c)=\sum_{J_k,J,N_k\leqslant N_c}\mathcal{M}_{\alpha}^{0\nu}(J_{k},J,N_k),
\end{eqnarray}
and in running nonclosure method as
\begin{eqnarray}
{M}_{\alpha}^{0\nu}(N_c)=\sum_{J_k,J,N_k\leqslant N_c} {M}_{\alpha}^{0\nu}(J_{k},J,N_k),
\end{eqnarray}
where $\mathcal{M}_{\alpha}^{0\nu}(J_{k},J,N_k)$ and ${M}_{\alpha}^{0\nu}(J_{k},J,N_k)$ is same as Eq. (\ref{eq:rcpartial}), and (\ref{eq:rncpartial}), respectively. 
Dependence of Fermi, Gamow-Teller, tensor, and total NMEs with $N_c$ is shown in Fig.~\ref{fig:nmevsnk}. Here, NMEs are calculated with total GXPF1A interaction in running closure and running nonclosure methods for AV18 SRC parametrization. The variation shows that the first few low lying states mostly contribute constructively and destructively. At a large value of $N_c$, NMEs becomes almost constant.
In Ref. \cite{PhysRevC.88.064312} it was found that considering 100 states for each
$J_k^{\pi}$ of intermediate nucleus $^{48}$Sc gives NMEs with less than 1\% uncertainty. In our calculation, we have considered $N_c$=150 for each allowed $J_k^{\pi}$ of $^{48}$Sc, which gives mostly constant NMEs. A similar dependence of NMEs with $N_c$ are seen for other SRC parametrization and for NMEs calculated with C and C+SO component of the GXPF1A interaction for different SRC parametrization. 

\begin{figure}
\centering
\includegraphics[trim=2cm 1cm 2cm 2cm,width=\linewidth]{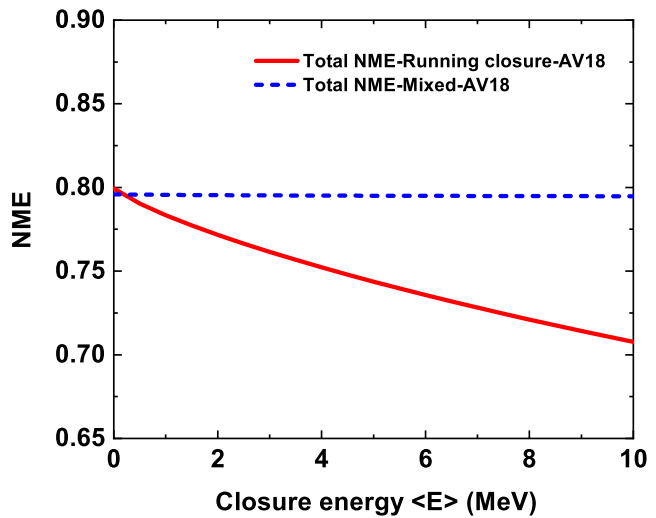}
\caption{\label{fig:nmevsclosure}(Color online) Dependence of NMEs for $0\nu\beta\beta$ (light neutrino exchange mechanism) of $^{48}$Ca with closure energy $\langle E\rangle$, calculated with total GXPF1A interaction for AV18 SRC parmaetrization in running closure and mixed methods.}
\end{figure}

Fig. \ref{fig:nmevsclosure} shows the variation of total NMEs in running closure and mixed methods with closure energy $\langle E\rangle$. NMEs shown here are calculated with total GXPF1A interaction for AV18 type SRC parametrization. It is found that for changing $\langle E\rangle$=0 to 10 MeV, there is about 11\% decrements of total NMEs in running closure method and almost negligible decrements of total NMEs in mixed method. Similar pattern of variation of NMEs with $\langle E\rangle$ are found with other SRC parametrization and in  NMEs calculated with C and C+SO components of the GXPF1A interaction for different SRC parametrization. 
\section{\label{sec:VI}Summary and Conclusions}
 In the present work, we have examined the role of C, SO, and T components GXPF1A effective interaction in NMEs for $2\nu\beta\beta$ and the light neutrino-exchange mechanism of $0\nu\beta\beta$ of $^{48}$Ca. The NMEs of $0\nu\beta\beta$ are calculated in closure, running closure, running nonclosure and mixed methods. The decomposition of the shell-model two-nucleon effective interaction into its individual components is performed using STD. 
 
 For $2\nu\beta\beta$, results show that the magnitude of NMEs calculated with the C+SO component of the interaction is about 7\% smaller than NMEs with C component of the interaction. NMEs further decrease by about 9\% in magnitude when NMEs are calculated by adding T component to the C+SO component of the interaction. 
 
 The NMEs of $0\nu\beta\beta$, calculated in running nonclosure method is enhanced by about 8-10\%, 8-10\%, and 9-12\% respectively as compared to corresponding NMEs in running closure method with C, C+SO, and total of GXPF1A interaction for different SRC parametrization. The NMEs for $2\nu\beta\beta$ and $0\nu\beta\beta$ have the same phase for C component and total GXPF1A interaction, and this phase is opposite as compared to C+SO component of the interaction. 
 
 We have also checked the contribution of each spin-parity ($J_k^{\pi}$) of the intermediate nucleus $^{48}$Sc in NMEs of $0\nu\beta\beta$ with running closure and running nonclosure method. It is found that for Fermi and Gamow-Teller NMEs, contribution of each $J_k^{\pi}$ is coherent with contribution through $1^{+}$, $3^{+}$, $5^{+}$ being dominating for Gamow-Teller NMEs and contribution through $2^{+}$, and $4^{+}$ being dominating for Fermi type NMEs. For tensor type NMEs, coherent contribution comes through 1$^{+}$, 3$^{+}$, 5$^{+}$, and 7$^{+}$ states with contribution from 3$^{+}$, and 5$^{+}$ being dominating. Contributions from 2$^{+}$, 4$^{+}$, and 6$^{+}$ states comes with the opposite phase and reduce the total tensor NMEs. The dominating enhancement of Gamow-Teller NMEs in running nonclosure method as compared to running closure method is found through 1$^{+}$ states.
 We have also examined the contributions in $0\nu\beta\beta$ NMEs in running closure and running closure method through different coupled spin-parity of two initial neutrons or final created protons ($J^\pi$). It is found that dominating contribution comes through 0$^{+}$ and 2$^{+}$ states with their phase being opposite, which reduces the total NMEs. 
 
Dependence of NMEs of $2\nu\beta\beta$ and $0\nu\beta\beta$ with cutoff number of states ($N_c$) and cutoff excitation energy ($E_c$) of intermediate nucleus ($^{48}$Sc) are also explored. It is found that only the first few low lying states contribute constructively and destructively in NMEs, and at large $N_c$ and $E_c$, NMEs becomes almost constant. This is because of the large value of neutrino momentum q ($\sim$ 100-200 MeV), whereas relevant excitation energy is only of the order of 10 MeV. In our case, we have considered $N_c=$150 for each $J_{k}^{\pi}$ of $^{48}$Sc with uncertainty being less than 1\%.

Variation of NMEs in closure and mixed methods with closure energy $\langle E\rangle$ are also examined. It is found that for changing $\langle E\rangle$=0 to 10 MeV, there is about 11\% decrements of total NMEs in running closure method and almost negligible decrements for total NMEs in mixed method.
\begin{acknowledgments}
S.S. thanks MHRD, Government of India for the providing fellowships for PhD.  
\end{acknowledgments}
\bibliography{main}

\begin{thebibliography}{53}%
\makeatletter
\providecommand \@ifxundefined [1]{%
 \@ifx{#1\undefined}
}%
\providecommand \@ifnum [1]{%
 \ifnum #1\expandafter \@firstoftwo
 \else \expandafter \@secondoftwo
 \fi
}%
\providecommand \@ifx [1]{%
 \ifx #1\expandafter \@firstoftwo
 \else \expandafter \@secondoftwo
 \fi
}%
\providecommand \natexlab [1]{#1}%
\providecommand \enquote  [1]{``#1''}%
\providecommand \bibnamefont  [1]{#1}%
\providecommand \bibfnamefont [1]{#1}%
\providecommand \citenamefont [1]{#1}%
\providecommand \href@noop [0]{\@secondoftwo}%
\providecommand \href [0]{\begingroup \@sanitize@url \@href}%
\providecommand \@href[1]{\@@startlink{#1}\@@href}%
\providecommand \@@href[1]{\endgroup#1\@@endlink}%
\providecommand \@sanitize@url [0]{\catcode `\\12\catcode `\$12\catcode
  `\&12\catcode `\#12\catcode `\^12\catcode `\_12\catcode `\%12\relax}%
\providecommand \@@startlink[1]{}%
\providecommand \@@endlink[0]{}%
\providecommand \url  [0]{\begingroup\@sanitize@url \@url }%
\providecommand \@url [1]{\endgroup\@href {#1}{\urlprefix }}%
\providecommand \urlprefix  [0]{URL }%
\providecommand \Eprint [0]{\href }%
\providecommand \doibase [0]{https://doi.org/}%
\providecommand \selectlanguage [0]{\@gobble}%
\providecommand \bibinfo  [0]{\@secondoftwo}%
\providecommand \bibfield  [0]{\@secondoftwo}%
\providecommand \translation [1]{[#1]}%
\providecommand \BibitemOpen [0]{}%
\providecommand \bibitemStop [0]{}%
\providecommand \bibitemNoStop [0]{.\EOS\space}%
\providecommand \EOS [0]{\spacefactor3000\relax}%
\providecommand \BibitemShut  [1]{\csname bibitem#1\endcsname}%
\let\auto@bib@innerbib\@empty
\bibitem [{\citenamefont {Goeppert-Mayer}(1935)}]{PhysRev.48.512}%
  \BibitemOpen
  \bibfield  {author} {\bibinfo {author} {\bibfnamefont {M.}~\bibnamefont
  {Goeppert-Mayer}},\ }\bibfield  {title} {\bibinfo {title} {Double
  beta-disintegration},\ }\href {https://doi.org/10.1103/PhysRev.48.512}
  {\bibfield  {journal} {\bibinfo  {journal} {Phys. Rev.}\ }\textbf {\bibinfo
  {volume} {48}},\ \bibinfo {pages} {512} (\bibinfo {year} {1935})}\BibitemShut
  {NoStop}%
\bibitem [{\citenamefont {Saakyan}(2013)}]{saakyan2013two}%
  \BibitemOpen
  \bibfield  {author} {\bibinfo {author} {\bibfnamefont {R.}~\bibnamefont
  {Saakyan}},\ }\bibfield  {title} {\bibinfo {title} {Two-neutrino double-beta
  decay},\ }\href@noop {} {\bibfield  {journal} {\bibinfo  {journal} {Annual
  Review of Nuclear and Particle Science}\ }\textbf {\bibinfo {volume} {63}},\
  \bibinfo {pages} {503} (\bibinfo {year} {2013})}\BibitemShut {NoStop}%
\bibitem [{\citenamefont {Furry}(1939)}]{PhysRev.56.1184}%
  \BibitemOpen
  \bibfield  {author} {\bibinfo {author} {\bibfnamefont {W.~H.}\ \bibnamefont
  {Furry}},\ }\bibfield  {title} {\bibinfo {title} {On transition probabilities
  in double beta-disintegration},\ }\href
  {https://doi.org/10.1103/PhysRev.56.1184} {\bibfield  {journal} {\bibinfo
  {journal} {Phys. Rev.}\ }\textbf {\bibinfo {volume} {56}},\ \bibinfo {pages}
  {1184} (\bibinfo {year} {1939})}\BibitemShut {NoStop}%
\bibitem [{\citenamefont {Vergados}\ \emph {et~al.}(2012)\citenamefont
  {Vergados}, \citenamefont {Ejiri},\ and\ \citenamefont
  {{\v{S}}imkovic}}]{vergados2012theory}%
  \BibitemOpen
  \bibfield  {author} {\bibinfo {author} {\bibfnamefont {J.~D.}\ \bibnamefont
  {Vergados}}, \bibinfo {author} {\bibfnamefont {H.}~\bibnamefont {Ejiri}},
  and\ \bibinfo {author} {\bibfnamefont {F.}~\bibnamefont {{\v{S}}imkovic}},\
  }\bibfield  {title} {\bibinfo {title} {Theory of neutrinoless double-beta
  decay},\ }\href@noop {} {\bibfield  {journal} {\bibinfo  {journal} {Reports
  on Progress in Physics}\ }\textbf {\bibinfo {volume} {75}},\ \bibinfo {pages}
  {106301} (\bibinfo {year} {2012})}\BibitemShut {NoStop}%
\bibitem [{\citenamefont {Majorana}(1937)}]{majorana1937teoria}%
  \BibitemOpen
  \bibfield  {author} {\bibinfo {author} {\bibfnamefont {E.}~\bibnamefont
  {Majorana}},\ }\bibfield  {title} {\bibinfo {title} {Symmetric theory of
  electron and positron},\ }\href@noop {} {\bibfield  {journal} {\bibinfo
  {journal} {Il Nuovo Cimento (1924-1942)}\ }\textbf {\bibinfo {volume} {14}},\
  \bibinfo {pages} {171} (\bibinfo {year} {1937})}\BibitemShut {NoStop}%
\bibitem [{\citenamefont {Racah}(1937)}]{racah1937sulla}%
  \BibitemOpen
  \bibfield  {author} {\bibinfo {author} {\bibfnamefont {G.}~\bibnamefont
  {Racah}},\ }\bibfield  {title} {\bibinfo {title} {Sulla simmetria tra
  particelle e antiparticelle},\ }\href@noop {} {\bibfield  {journal} {\bibinfo
   {journal} {Il Nuovo Cimento}\ }\textbf {\bibinfo {volume} {14}},\ \bibinfo
  {pages} {322} (\bibinfo {year} {1937})}\BibitemShut {NoStop}%
\bibitem [{\citenamefont {Schechter$~$}\ and\ \citenamefont
  {Valle}(1982)}]{PhysRevD.25.2951}%
  \BibitemOpen
  \bibfield  {author} {\bibinfo {author} {\bibfnamefont {J.}~\bibnamefont
  {Schechter$~$}}and\ \bibinfo {author} {\bibfnamefont {J.~W.}\ \bibnamefont
  {Valle}},\ }\bibfield  {title} {\bibinfo {title} {Neutrinoless double-$\beta$
  decay in su (2)$\times$ u (1) theories},\ }\href@noop {} {\bibfield
  {journal} {\bibinfo  {journal} {Physical Review D}\ }\textbf {\bibinfo
  {volume} {25}},\ \bibinfo {pages} {2951} (\bibinfo {year}
  {1982})}\BibitemShut {NoStop}%
\bibitem [{\citenamefont {Deppisch}\ \emph {et~al.}(2012)\citenamefont
  {Deppisch}, \citenamefont {Hirsch},\ and\ \citenamefont
  {P{\"a}s}}]{deppisch2012neutrinoless}%
  \BibitemOpen
  \bibfield  {author} {\bibinfo {author} {\bibfnamefont {F.~F.}\ \bibnamefont
  {Deppisch}}, \bibinfo {author} {\bibfnamefont {M.}~\bibnamefont {Hirsch}},
  and\ \bibinfo {author} {\bibfnamefont {H.}~\bibnamefont {P{\"a}s}},\
  }\bibfield  {title} {\bibinfo {title} {Neutrinoless double-beta decay and
  physics beyond the standard model},\ }\href@noop {} {\bibfield  {journal}
  {\bibinfo  {journal} {Journal of Physics G: Nuclear and Particle Physics}\
  }\textbf {\bibinfo {volume} {39}},\ \bibinfo {pages} {124007} (\bibinfo
  {year} {2012})}\BibitemShut {NoStop}%
\bibitem [{\citenamefont {Rodejohann}(2011)}]{rodejohann2011neutrino}%
  \BibitemOpen
  \bibfield  {author} {\bibinfo {author} {\bibfnamefont {W.}~\bibnamefont
  {Rodejohann}},\ }\bibfield  {title} {\bibinfo {title} {Neutrino-less double
  beta decay and particle physics},\ }\href@noop {} {\bibfield  {journal}
  {\bibinfo  {journal} {International Journal of Modern Physics E}\ }\textbf
  {\bibinfo {volume} {20}},\ \bibinfo {pages} {1833} (\bibinfo {year}
  {2011})}\BibitemShut {NoStop}%
\bibitem [{\citenamefont {Avignone}\ \emph {et~al.}(2008)\citenamefont
  {Avignone}, \citenamefont {Elliott},\ and\ \citenamefont
  {Engel}}]{RevModPhys.80.481}%
  \BibitemOpen
  \bibfield  {author} {\bibinfo {author} {\bibfnamefont {F.~T.}\ \bibnamefont
  {Avignone}}, \bibinfo {author} {\bibfnamefont {S.~R.}\ \bibnamefont
  {Elliott}}, and\ \bibinfo {author} {\bibfnamefont {J.}~\bibnamefont
  {Engel}},\ }\bibfield  {title} {\bibinfo {title} {Double beta decay, majorana
  neutrinos, and neutrino mass},\ }\href
  {https://doi.org/10.1103/RevModPhys.80.481} {\bibfield  {journal} {\bibinfo
  {journal} {Rev. Mod. Phys.}\ }\textbf {\bibinfo {volume} {80}},\ \bibinfo
  {pages} {481} (\bibinfo {year} {2008})}\BibitemShut {NoStop}%
\bibitem [{\citenamefont {Engel$~$}\ and\ \citenamefont
  {Men{\'e}ndez}(2017)}]{engel2017status}%
  \BibitemOpen
  \bibfield  {author} {\bibinfo {author} {\bibfnamefont {J.}~\bibnamefont
  {Engel$~$}}and\ \bibinfo {author} {\bibfnamefont {J.}~\bibnamefont
  {Men{\'e}ndez}},\ }\bibfield  {title} {\bibinfo {title} {Status and future of
  nuclear matrix elements for neutrinoless double-beta decay: a review},\
  }\href@noop {} {\bibfield  {journal} {\bibinfo  {journal} {Reports on
  Progress in Physics}\ }\textbf {\bibinfo {volume} {80}},\ \bibinfo {pages}
  {046301} (\bibinfo {year} {2017})}\BibitemShut {NoStop}%
\bibitem [{\citenamefont {Caurier}\ \emph {et~al.}(2008)\citenamefont
  {Caurier}, \citenamefont {Men\'endez}, \citenamefont {Nowacki},\ and\
  \citenamefont {Poves}}]{PhysRevLett.100.052503}%
  \BibitemOpen
  \bibfield  {author} {\bibinfo {author} {\bibfnamefont {E.}~\bibnamefont
  {Caurier}}, \bibinfo {author} {\bibfnamefont {J.}~\bibnamefont {Men\'endez}},
  \bibinfo {author} {\bibfnamefont {F.}~\bibnamefont {Nowacki}}, and\ \bibinfo
  {author} {\bibfnamefont {A.}~\bibnamefont {Poves}},\ }\bibfield  {title}
  {\bibinfo {title} {Influence of pairing on the nuclear matrix elements of the
  neutrinoless $\ensuremath{\beta}\ensuremath{\beta}$ decays},\ }\href
  {https://doi.org/10.1103/PhysRevLett.100.052503} {\bibfield  {journal}
  {\bibinfo  {journal} {Phys. Rev. Lett.}\ }\textbf {\bibinfo {volume} {100}},\
  \bibinfo {pages} {052503} (\bibinfo {year} {2008})}\BibitemShut {NoStop}%
\bibitem [{\citenamefont {Horoi}\ and\ \citenamefont
  {Stoica}(2010)}]{PhysRevC.81.024321}%
  \BibitemOpen
  \bibfield  {author} {\bibinfo {author} {\bibfnamefont {M.}~\bibnamefont
  {Horoi}}and\ \bibinfo {author} {\bibfnamefont {S.}~\bibnamefont {Stoica}},\
  }\bibfield  {title} {\bibinfo {title} {Shell model analysis of the
  neutrinoless double-$\ensuremath{\beta}$ decay of $^{48}\mathrm{Ca}$},\
  }\href {https://doi.org/10.1103/PhysRevC.81.024321} {\bibfield  {journal}
  {\bibinfo  {journal} {Phys. Rev. C}\ }\textbf {\bibinfo {volume} {81}},\
  \bibinfo {pages} {024321} (\bibinfo {year} {2010})}\BibitemShut {NoStop}%
\bibitem [{\citenamefont {Sen'kov$~$}\ and\ \citenamefont
  {Horoi}(2013)}]{PhysRevC.88.064312}%
  \BibitemOpen
  \bibfield  {author} {\bibinfo {author} {\bibfnamefont {R.~A.}\ \bibnamefont
  {Sen'kov$~$}}and\ \bibinfo {author} {\bibfnamefont {M.}~\bibnamefont
  {Horoi}},\ }\bibfield  {title} {\bibinfo {title} {Neutrinoless
  double-$\ensuremath{\beta}$ decay of ${}^{48}$ca in the shell model: Closure
  versus nonclosure approximation},\ }\href
  {https://doi.org/10.1103/PhysRevC.88.064312} {\bibfield  {journal} {\bibinfo
  {journal} {Phys. Rev. C}\ }\textbf {\bibinfo {volume} {88}},\ \bibinfo
  {pages} {064312} (\bibinfo {year} {2013})}\BibitemShut {NoStop}%
\bibitem [{\citenamefont {Brown}\ \emph {et~al.}(2014)\citenamefont {Brown},
  \citenamefont {Horoi},\ and\ \citenamefont
  {Sen'kov}}]{PhysRevLett.113.262501}%
  \BibitemOpen
  \bibfield  {author} {\bibinfo {author} {\bibfnamefont {B.~A.}\ \bibnamefont
  {Brown}}, \bibinfo {author} {\bibfnamefont {M.}~\bibnamefont {Horoi}}, and\
  \bibinfo {author} {\bibfnamefont {R.~A.}\ \bibnamefont {Sen'kov}},\
  }\bibfield  {title} {\bibinfo {title} {Nuclear structure aspects of
  neutrinoless double-$\ensuremath{\beta}$ decay},\ }\href
  {https://doi.org/10.1103/PhysRevLett.113.262501} {\bibfield  {journal}
  {\bibinfo  {journal} {Phys. Rev. Lett.}\ }\textbf {\bibinfo {volume} {113}},\
  \bibinfo {pages} {262501} (\bibinfo {year} {2014})}\BibitemShut {NoStop}%
\bibitem [{\citenamefont {Iwata}\ \emph {et~al.}(2016)\citenamefont {Iwata},
  \citenamefont {Shimizu}, \citenamefont {Otsuka}, \citenamefont {Utsuno},
  \citenamefont {Men\'endez}, \citenamefont {Honma},\ and\ \citenamefont
  {Abe}}]{PhysRevLett.116.112502}%
  \BibitemOpen
  \bibfield  {author} {\bibinfo {author} {\bibfnamefont {Y.}~\bibnamefont
  {Iwata}}, \bibinfo {author} {\bibfnamefont {N.}~\bibnamefont {Shimizu}},
  \bibinfo {author} {\bibfnamefont {T.}~\bibnamefont {Otsuka}}, \bibinfo
  {author} {\bibfnamefont {Y.}~\bibnamefont {Utsuno}}, \bibinfo {author}
  {\bibfnamefont {J.}~\bibnamefont {Men\'endez}}, \bibinfo {author}
  {\bibfnamefont {M.}~\bibnamefont {Honma}}, and\ \bibinfo {author}
  {\bibfnamefont {T.}~\bibnamefont {Abe}},\ }\bibfield  {title} {\bibinfo
  {title} {Large-scale shell-model analysis of the neutrinoless
  $\ensuremath{\beta}\ensuremath{\beta}$ decay of $^{48}\mathrm{Ca}$},\ }\href
  {https://doi.org/10.1103/PhysRevLett.116.112502} {\bibfield  {journal}
  {\bibinfo  {journal} {Phys. Rev. Lett.}\ }\textbf {\bibinfo {volume} {116}},\
  \bibinfo {pages} {112502} (\bibinfo {year} {2016})}\BibitemShut {NoStop}%
\bibitem [{\citenamefont {Barea$~$}\ and\ \citenamefont
  {Iachello}(2009)}]{PhysRevC.79.044301}%
  \BibitemOpen
  \bibfield  {author} {\bibinfo {author} {\bibfnamefont {J.}~\bibnamefont
  {Barea$~$}}and\ \bibinfo {author} {\bibfnamefont {F.}~\bibnamefont
  {Iachello}},\ }\bibfield  {title} {\bibinfo {title} {Neutrinoless
  double-$\ensuremath{\beta}$ decay in the microscopic interacting boson
  model},\ }\href {https://doi.org/10.1103/PhysRevC.79.044301} {\bibfield
  {journal} {\bibinfo  {journal} {Phys. Rev. C}\ }\textbf {\bibinfo {volume}
  {79}},\ \bibinfo {pages} {044301} (\bibinfo {year} {2009})}\BibitemShut
  {NoStop}%
\bibitem [{\citenamefont {Barea}\ \emph {et~al.}(2012)\citenamefont {Barea},
  \citenamefont {Kotila},\ and\ \citenamefont
  {Iachello}}]{PhysRevLett.109.042501}%
  \BibitemOpen
  \bibfield  {author} {\bibinfo {author} {\bibfnamefont {J.}~\bibnamefont
  {Barea}}, \bibinfo {author} {\bibfnamefont {J.}~\bibnamefont {Kotila}}, and\
  \bibinfo {author} {\bibfnamefont {F.}~\bibnamefont {Iachello}},\ }\bibfield
  {title} {\bibinfo {title} {Limits on neutrino masses from neutrinoless
  double-$\ensuremath{\beta}$ decay},\ }\href
  {https://doi.org/10.1103/PhysRevLett.109.042501} {\bibfield  {journal}
  {\bibinfo  {journal} {Phys. Rev. Lett.}\ }\textbf {\bibinfo {volume} {109}},\
  \bibinfo {pages} {042501} (\bibinfo {year} {2012})}\BibitemShut {NoStop}%
\bibitem [{\citenamefont {Rodr\'{\i}guez$~$}\ and\ \citenamefont
  {Mart\'{\i}nez-Pinedo}(2010)}]{PhysRevLett.105.252503}%
  \BibitemOpen
  \bibfield  {author} {\bibinfo {author} {\bibfnamefont {T.~R.}\ \bibnamefont
  {Rodr\'{\i}guez$~$}}and\ \bibinfo {author} {\bibfnamefont {G.}~\bibnamefont
  {Mart\'{\i}nez-Pinedo}},\ }\bibfield  {title} {\bibinfo {title} {Energy
  density functional study of nuclear matrix elements for neutrinoless
  $\ensuremath{\beta}\ensuremath{\beta}$ decay},\ }\href
  {https://doi.org/10.1103/PhysRevLett.105.252503} {\bibfield  {journal}
  {\bibinfo  {journal} {Phys. Rev. Lett.}\ }\textbf {\bibinfo {volume} {105}},\
  \bibinfo {pages} {252503} (\bibinfo {year} {2010})}\BibitemShut {NoStop}%
\bibitem [{\citenamefont {Song}\ \emph {et~al.}(2014)\citenamefont {Song},
  \citenamefont {Yao}, \citenamefont {Ring},\ and\ \citenamefont
  {Meng}}]{PhysRevC.90.054309}%
  \BibitemOpen
  \bibfield  {author} {\bibinfo {author} {\bibfnamefont {L.~S.}\ \bibnamefont
  {Song}}, \bibinfo {author} {\bibfnamefont {J.~M.}\ \bibnamefont {Yao}},
  \bibinfo {author} {\bibfnamefont {P.}~\bibnamefont {Ring}}, and\ \bibinfo
  {author} {\bibfnamefont {J.}~\bibnamefont {Meng}},\ }\bibfield  {title}
  {\bibinfo {title} {Relativistic description of nuclear matrix elements in
  neutrinoless double-$\ensuremath{\beta}$ decay},\ }\href
  {https://doi.org/10.1103/PhysRevC.90.054309} {\bibfield  {journal} {\bibinfo
  {journal} {Phys. Rev. C}\ }\textbf {\bibinfo {volume} {90}},\ \bibinfo
  {pages} {054309} (\bibinfo {year} {2014})}\BibitemShut {NoStop}%
\bibitem [{\citenamefont {Rath}\ \emph {et~al.}(2010)\citenamefont {Rath},
  \citenamefont {Chandra}, \citenamefont {Chaturvedi}, \citenamefont {Raina},\
  and\ \citenamefont {Hirsch}}]{PhysRevC.82.064310}%
  \BibitemOpen
  \bibfield  {author} {\bibinfo {author} {\bibfnamefont {P.~K.}\ \bibnamefont
  {Rath}}, \bibinfo {author} {\bibfnamefont {R.}~\bibnamefont {Chandra}},
  \bibinfo {author} {\bibfnamefont {K.}~\bibnamefont {Chaturvedi}}, \bibinfo
  {author} {\bibfnamefont {P.~K.}\ \bibnamefont {Raina}}, and\ \bibinfo
  {author} {\bibfnamefont {J.~G.}\ \bibnamefont {Hirsch}},\ }\bibfield  {title}
  {\bibinfo {title} {Uncertainties in nuclear transition matrix elements for
  neutrinoless $\ensuremath{\beta}\ensuremath{\beta}$ decay within the
  projected-hartree-fock-bogoliubov model},\ }\href
  {https://doi.org/10.1103/PhysRevC.82.064310} {\bibfield  {journal} {\bibinfo
  {journal} {Phys. Rev. C}\ }\textbf {\bibinfo {volume} {82}},\ \bibinfo
  {pages} {064310} (\bibinfo {year} {2010})}\BibitemShut {NoStop}%
\bibitem [{\citenamefont {Rodin}\ \emph {et~al.}(2006)\citenamefont {Rodin},
  \citenamefont {Faessler}, \citenamefont {{\v{S}}imkovic},\ and\ \citenamefont
  {Vogel}}]{rodin2006assessment}%
  \BibitemOpen
  \bibfield  {author} {\bibinfo {author} {\bibfnamefont {V.}~\bibnamefont
  {Rodin}}, \bibinfo {author} {\bibfnamefont {A.}~\bibnamefont {Faessler}},
  \bibinfo {author} {\bibfnamefont {F.}~\bibnamefont {{\v{S}}imkovic}}, and\
  \bibinfo {author} {\bibfnamefont {P.}~\bibnamefont {Vogel}},\ }\bibfield
  {title} {\bibinfo {title} {Assessment of uncertainties in qrpa
  0$\nu$$\beta$$\beta$-decay nuclear matrix elements},\ }\href@noop {}
  {\bibfield  {journal} {\bibinfo  {journal} {Nuclear Physics A}\ }\textbf
  {\bibinfo {volume} {766}},\ \bibinfo {pages} {107} (\bibinfo {year}
  {2006})}\BibitemShut {NoStop}%
\bibitem [{\citenamefont {\ifmmode~\check{S}\else \v{S}\fi{}imkovic}\ \emph
  {et~al.}(1999)\citenamefont {\ifmmode~\check{S}\else \v{S}\fi{}imkovic},
  \citenamefont {Pantis}, \citenamefont {Vergados},\ and\ \citenamefont
  {Faessler}}]{PhysRevC.60.055502}%
  \BibitemOpen
  \bibfield  {author} {\bibinfo {author} {\bibfnamefont {F.}~\bibnamefont
  {\ifmmode~\check{S}\else \v{S}\fi{}imkovic}}, \bibinfo {author}
  {\bibfnamefont {G.}~\bibnamefont {Pantis}}, \bibinfo {author} {\bibfnamefont
  {J.~D.}\ \bibnamefont {Vergados}}, and\ \bibinfo {author} {\bibfnamefont
  {A.}~\bibnamefont {Faessler}},\ }\bibfield  {title} {\bibinfo {title}
  {Additional nucleon current contributions to neutrinoless double
  $\ensuremath{\beta}$ decay},\ }\href
  {https://doi.org/10.1103/PhysRevC.60.055502} {\bibfield  {journal} {\bibinfo
  {journal} {Phys. Rev. C}\ }\textbf {\bibinfo {volume} {60}},\ \bibinfo
  {pages} {055502} (\bibinfo {year} {1999})}\BibitemShut {NoStop}%
\bibitem [{\citenamefont {Mohapatra$~$}\ and\ \citenamefont
  {Senjanovi\ifmmode~\acute{c}\else \'{c}\fi{}}(1980)}]{PhysRevLett.44.912}%
  \BibitemOpen
  \bibfield  {author} {\bibinfo {author} {\bibfnamefont {R.~N.}\ \bibnamefont
  {Mohapatra$~$}}and\ \bibinfo {author} {\bibfnamefont {G.}~\bibnamefont
  {Senjanovi\ifmmode~\acute{c}\else \'{c}\fi{}}},\ }\bibfield  {title}
  {\bibinfo {title} {Neutrino mass and spontaneous parity nonconservation},\
  }\href {https://doi.org/10.1103/PhysRevLett.44.912} {\bibfield  {journal}
  {\bibinfo  {journal} {Phys. Rev. Lett.}\ }\textbf {\bibinfo {volume} {44}},\
  \bibinfo {pages} {912} (\bibinfo {year} {1980})}\BibitemShut {NoStop}%
\bibitem [{\citenamefont {Mohapatra$~$}\ and\ \citenamefont
  {Vergados}(1981)}]{PhysRevLett.47.1713}%
  \BibitemOpen
  \bibfield  {author} {\bibinfo {author} {\bibfnamefont {R.~N.}\ \bibnamefont
  {Mohapatra$~$}}and\ \bibinfo {author} {\bibfnamefont {J.~D.}\ \bibnamefont
  {Vergados}},\ }\bibfield  {title} {\bibinfo {title} {New contribution to
  neutrinoless double beta decay in gauge models},\ }\href
  {https://doi.org/10.1103/PhysRevLett.47.1713} {\bibfield  {journal} {\bibinfo
   {journal} {Phys. Rev. Lett.}\ }\textbf {\bibinfo {volume} {47}},\ \bibinfo
  {pages} {1713} (\bibinfo {year} {1981})}\BibitemShut {NoStop}%
\bibitem [{\citenamefont {Mohapatra}(1986)}]{PhysRevD.34.3457}%
  \BibitemOpen
  \bibfield  {author} {\bibinfo {author} {\bibfnamefont {R.~N.}\ \bibnamefont
  {Mohapatra}},\ }\bibfield  {title} {\bibinfo {title} {New contributions to
  neutrinoless double-beta decay in supersymmetric theories},\ }\href
  {https://doi.org/10.1103/PhysRevD.34.3457} {\bibfield  {journal} {\bibinfo
  {journal} {Phys. Rev. D}\ }\textbf {\bibinfo {volume} {34}},\ \bibinfo
  {pages} {3457} (\bibinfo {year} {1986})}\BibitemShut {NoStop}%
\bibitem [{\citenamefont {Vergados}(1987)}]{vergados1987neutrinoless}%
  \BibitemOpen
  \bibfield  {author} {\bibinfo {author} {\bibfnamefont {J.}~\bibnamefont
  {Vergados}},\ }\bibfield  {title} {\bibinfo {title} {Neutrinoless double
  $\beta$-decay without majorana neutrinos in supersymmetric theories},\
  }\href@noop {} {\bibfield  {journal} {\bibinfo  {journal} {Physics Letters
  B}\ }\textbf {\bibinfo {volume} {184}},\ \bibinfo {pages} {55} (\bibinfo
  {year} {1987})}\BibitemShut {NoStop}%
\bibitem [{\citenamefont {Horoi}\ \emph {et~al.}(2007)\citenamefont {Horoi},
  \citenamefont {Stoica},\ and\ \citenamefont {Brown}}]{PhysRevC.75.034303}%
  \BibitemOpen
  \bibfield  {author} {\bibinfo {author} {\bibfnamefont {M.}~\bibnamefont
  {Horoi}}, \bibinfo {author} {\bibfnamefont {S.}~\bibnamefont {Stoica}}, and\
  \bibinfo {author} {\bibfnamefont {B.~A.}\ \bibnamefont {Brown}},\ }\bibfield
  {title} {\bibinfo {title} {Shell-model calculations of two-neutrino
  double-\ensuremath{\beta} decay rates of $^{48}\mathrm{Ca}$ with the gxpf1a
  interaction},\ }\href {https://doi.org/10.1103/PhysRevC.75.034303} {\bibfield
   {journal} {\bibinfo  {journal} {Phys. Rev. C}\ }\textbf {\bibinfo {volume}
  {75}},\ \bibinfo {pages} {034303} (\bibinfo {year} {2007})}\BibitemShut
  {NoStop}%
\bibitem [{\citenamefont {Zhao}\ \emph {et~al.}(1990)\citenamefont {Zhao},
  \citenamefont {Brown},\ and\ \citenamefont {Richter}}]{PhysRevC.42.1120}%
  \BibitemOpen
  \bibfield  {author} {\bibinfo {author} {\bibfnamefont {L.}~\bibnamefont
  {Zhao}}, \bibinfo {author} {\bibfnamefont {B.~A.}\ \bibnamefont {Brown}},
  and\ \bibinfo {author} {\bibfnamefont {W.~A.}\ \bibnamefont {Richter}},\
  }\bibfield  {title} {\bibinfo {title} {Shell-model calculation for
  two-neutrino double beta decay of $^{48}\mathrm{Ca}$},\ }\href
  {https://doi.org/10.1103/PhysRevC.42.1120} {\bibfield  {journal} {\bibinfo
  {journal} {Phys. Rev. C}\ }\textbf {\bibinfo {volume} {42}},\ \bibinfo
  {pages} {1120} (\bibinfo {year} {1990})}\BibitemShut {NoStop}%
\bibitem [{\citenamefont {Caurier}\ \emph {et~al.}(1990)\citenamefont
  {Caurier}, \citenamefont {Poves},\ and\ \citenamefont
  {Zuker}}]{caurier1990full}%
  \BibitemOpen
  \bibfield  {author} {\bibinfo {author} {\bibfnamefont {E.}~\bibnamefont
  {Caurier}}, \bibinfo {author} {\bibfnamefont {A.}~\bibnamefont {Poves}}, and\
  \bibinfo {author} {\bibfnamefont {A.}~\bibnamefont {Zuker}},\ }\bibfield
  {title} {\bibinfo {title} {A full 0h$\omega$ description of the
  2$\nu$$\beta$$\beta$ decay of 48ca},\ }\href@noop {} {\bibfield  {journal}
  {\bibinfo  {journal} {Physics Letters B}\ }\textbf {\bibinfo {volume}
  {252}},\ \bibinfo {pages} {13} (\bibinfo {year} {1990})}\BibitemShut
  {NoStop}%
\bibitem [{\citenamefont {Kumar}\ \emph
  {et~al.}(2019{\natexlab{a}})\citenamefont {Kumar}, \citenamefont {Jha},
  \citenamefont {K.~Raina},\ and\ \citenamefont {Singh}}]{kumar2019quasi}%
  \BibitemOpen
  \bibfield  {author} {\bibinfo {author} {\bibfnamefont {P.}~\bibnamefont
  {Kumar}}, \bibinfo {author} {\bibfnamefont {K.}~\bibnamefont {Jha}}, \bibinfo
  {author} {\bibfnamefont {P.}~\bibnamefont {K.~Raina}}, and\ \bibinfo {author}
  {\bibfnamefont {P.~P.}\ \bibnamefont {Singh}},\ }\bibfield  {title} {\bibinfo
  {title} {Quasi shell gap at $^{23}\mathrm{F}$},\ }\href@noop {} {\bibfield
  {journal} {\bibinfo  {journal} {Nuclear Physics A}\ }\textbf {\bibinfo
  {volume} {983}},\ \bibinfo {pages} {210} (\bibinfo {year}
  {2019}{\natexlab{a}})}\BibitemShut {NoStop}%
\bibitem [{\citenamefont {Kumar}\ \emph
  {et~al.}(2019{\natexlab{b}})\citenamefont {Kumar}, \citenamefont {Sarkar},
  \citenamefont {Singh},\ and\ \citenamefont {Raina}}]{PhysRevC.100.024328}%
  \BibitemOpen
  \bibfield  {author} {\bibinfo {author} {\bibfnamefont {P.}~\bibnamefont
  {Kumar}}, \bibinfo {author} {\bibfnamefont {S.}~\bibnamefont {Sarkar}},
  \bibinfo {author} {\bibfnamefont {P.~P.}\ \bibnamefont {Singh}}, and\
  \bibinfo {author} {\bibfnamefont {P.~K.}\ \bibnamefont {Raina}},\ }\bibfield
  {title} {\bibinfo {title} {Proton-neutron force and proton single-particle
  strength in $\mathrm{Sc}$, $\mathrm{F}$, and $\mathrm{Li}$ isotopes},\ }\href
  {https://doi.org/10.1103/PhysRevC.100.024328} {\bibfield  {journal} {\bibinfo
   {journal} {Phys. Rev. C}\ }\textbf {\bibinfo {volume} {100}},\ \bibinfo
  {pages} {024328} (\bibinfo {year} {2019}{\natexlab{b}})}\BibitemShut
  {NoStop}%
\bibitem [{\citenamefont {Umeya$~$}\ and\ \citenamefont
  {Muto}(2004)}]{PhysRevC.69.024306}%
  \BibitemOpen
  \bibfield  {author} {\bibinfo {author} {\bibfnamefont {A.}~\bibnamefont
  {Umeya$~$}}and\ \bibinfo {author} {\bibfnamefont {K.}~\bibnamefont {Muto}},\
  }\bibfield  {title} {\bibinfo {title} {Triplet-even channel attraction for
  shell gaps},\ }\href {https://doi.org/10.1103/PhysRevC.69.024306} {\bibfield
  {journal} {\bibinfo  {journal} {Phys. Rev. C}\ }\textbf {\bibinfo {volume}
  {69}},\ \bibinfo {pages} {024306} (\bibinfo {year} {2004})}\BibitemShut
  {NoStop}%
\bibitem [{\citenamefont {Umeya$~$}\ and\ \citenamefont
  {Muto}(2006)}]{PhysRevC.74.034330}%
  \BibitemOpen
  \bibfield  {author} {\bibinfo {author} {\bibfnamefont {A.}~\bibnamefont
  {Umeya$~$}}and\ \bibinfo {author} {\bibfnamefont {K.}~\bibnamefont {Muto}},\
  }\bibfield  {title} {\bibinfo {title} {Single-particle energies in
  neutron-rich nuclei by shell model sum rule},\ }\href
  {https://doi.org/10.1103/PhysRevC.74.034330} {\bibfield  {journal} {\bibinfo
  {journal} {Phys. Rev. C}\ }\textbf {\bibinfo {volume} {74}},\ \bibinfo
  {pages} {034330} (\bibinfo {year} {2006})}\BibitemShut {NoStop}%
\bibitem [{\citenamefont {Smirnova}\ \emph {et~al.}(2012)\citenamefont
  {Smirnova}, \citenamefont {Heyde}, \citenamefont {Bally}, \citenamefont
  {Nowacki},\ and\ \citenamefont {Sieja}}]{PhysRevC.86.034314}%
  \BibitemOpen
  \bibfield  {author} {\bibinfo {author} {\bibfnamefont {N.~A.}\ \bibnamefont
  {Smirnova}}, \bibinfo {author} {\bibfnamefont {K.}~\bibnamefont {Heyde}},
  \bibinfo {author} {\bibfnamefont {B.}~\bibnamefont {Bally}}, \bibinfo
  {author} {\bibfnamefont {F.}~\bibnamefont {Nowacki}}, and\ \bibinfo {author}
  {\bibfnamefont {K.}~\bibnamefont {Sieja}},\ }\bibfield  {title} {\bibinfo
  {title} {Nuclear shell evolution and in-medium $nn$ interaction},\ }\href
  {https://doi.org/10.1103/PhysRevC.86.034314} {\bibfield  {journal} {\bibinfo
  {journal} {Phys. Rev. C}\ }\textbf {\bibinfo {volume} {86}},\ \bibinfo
  {pages} {034314} (\bibinfo {year} {2012})}\BibitemShut {NoStop}%
\bibitem [{\citenamefont {Sarkar}\ \emph {et~al.}(2020)\citenamefont {Sarkar},
  \citenamefont {Kumar}, \citenamefont {Jha},\ and\ \citenamefont
  {Raina}}]{PhysRevC.101.014307}%
  \BibitemOpen
  \bibfield  {author} {\bibinfo {author} {\bibfnamefont {S.}~\bibnamefont
  {Sarkar}}, \bibinfo {author} {\bibfnamefont {P.}~\bibnamefont {Kumar}},
  \bibinfo {author} {\bibfnamefont {K.}~\bibnamefont {Jha}}, and\ \bibinfo
  {author} {\bibfnamefont {P.~K.}\ \bibnamefont {Raina}},\ }\bibfield  {title}
  {\bibinfo {title} {Sensitivity of nuclear matrix elements of
  $0\ensuremath{\nu}\ensuremath{\beta}\ensuremath{\beta}$ of $^{48}\mathrm{Ca}$
  to different components of the two-nucleon interaction},\ }\href
  {https://doi.org/10.1103/PhysRevC.101.014307} {\bibfield  {journal} {\bibinfo
   {journal} {Phys. Rev. C}\ }\textbf {\bibinfo {volume} {101}},\ \bibinfo
  {pages} {014307} (\bibinfo {year} {2020})}\BibitemShut {NoStop}%
\bibitem [{\citenamefont {P.~Elliott~\textit{et al.}}(1968)}]{dirim1968jp}%
  \BibitemOpen
  \bibfield  {author} {\bibinfo {author} {\bibfnamefont {J.}~\bibnamefont
  {P.~Elliott~\textit{et al.}}},\ }\bibfield  {title} {\bibinfo {title} {Matrix
  elements of the nucleon-nucleon potential for use in nuclear-structure
  calculations},\ }\href@noop {} {\bibfield  {journal} {\bibinfo  {journal}
  {Nucl. Phys}\ }\textbf {\bibinfo {volume} {121}},\ \bibinfo {pages} {241}
  (\bibinfo {year} {1968})}\BibitemShut {NoStop}%
\bibitem [{\citenamefont {Kirson}(1973)}]{Kirson:1973ffz}%
  \BibitemOpen
  \bibfield  {author} {\bibinfo {author} {\bibfnamefont {M.~W.}\ \bibnamefont
  {Kirson}},\ }\bibfield  {title} {\bibinfo {title} {{Spin-tensor decomposition
  of nuclear effective interactions}},\ }\href
  {https://doi.org/10.1016/0370-2693(73)90582-0} {\bibfield  {journal}
  {\bibinfo  {journal} {Phys. Lett.}\ }\textbf {\bibinfo {volume} {47B}},\
  \bibinfo {pages} {110} (\bibinfo {year} {1973})}\BibitemShut {NoStop}%
\bibitem [{\citenamefont {Schiffer$~$}\ and\ \citenamefont
  {True}(1976)}]{RevModPhys.48.191}%
  \BibitemOpen
  \bibfield  {author} {\bibinfo {author} {\bibfnamefont {J.~P.}\ \bibnamefont
  {Schiffer$~$}}and\ \bibinfo {author} {\bibfnamefont {W.~W.}\ \bibnamefont
  {True}},\ }\bibfield  {title} {\bibinfo {title} {The effective interaction
  between nucleons deduced from nuclear spectra},\ }\href
  {https://doi.org/10.1103/RevModPhys.48.191} {\bibfield  {journal} {\bibinfo
  {journal} {Rev. Mod. Phys.}\ }\textbf {\bibinfo {volume} {48}},\ \bibinfo
  {pages} {191} (\bibinfo {year} {1976})}\BibitemShut {NoStop}%
\bibitem [{\citenamefont {Klingenbeck}\ \emph {et~al.}(1977)\citenamefont
  {Klingenbeck}, \citenamefont {Kn\"upfer}, \citenamefont {Huber},\ and\
  \citenamefont {Glaudemans}}]{PhysRevC.15.1483}%
  \BibitemOpen
  \bibfield  {author} {\bibinfo {author} {\bibfnamefont {K.}~\bibnamefont
  {Klingenbeck}}, \bibinfo {author} {\bibfnamefont {W.}~\bibnamefont
  {Kn\"upfer}}, \bibinfo {author} {\bibfnamefont {M.~G.}\ \bibnamefont
  {Huber}}, and\ \bibinfo {author} {\bibfnamefont {P.~W.~M.}\ \bibnamefont
  {Glaudemans}},\ }\bibfield  {title} {\bibinfo {title} {Central and noncentral
  components of the effective $\mathrm{sd}$-shell interaction},\ }\href
  {https://doi.org/10.1103/PhysRevC.15.1483} {\bibfield  {journal} {\bibinfo
  {journal} {Phys. Rev. C}\ }\textbf {\bibinfo {volume} {15}},\ \bibinfo
  {pages} {1483} (\bibinfo {year} {1977})}\BibitemShut {NoStop}%
\bibitem [{\citenamefont {Kenji}(1980)}]{kenji1980spin}%
  \BibitemOpen
  \bibfield  {author} {\bibinfo {author} {\bibfnamefont {Y.}~\bibnamefont
  {Kenji}},\ }\bibfield  {title} {\bibinfo {title} {Spin-tensor decomposition
  of effective interactions for 0p shell nuclei},\ }\href@noop {} {\bibfield
  {journal} {\bibinfo  {journal} {Nuclear Physics A}\ }\textbf {\bibinfo
  {volume} {333}},\ \bibinfo {pages} {67} (\bibinfo {year} {1980})}\BibitemShut
  {NoStop}%
\bibitem [{\citenamefont {Smirnova}\ \emph {et~al.}(2010)\citenamefont
  {Smirnova}, \citenamefont {Bally}, \citenamefont {Heyde}, \citenamefont
  {Nowacki},\ and\ \citenamefont {Sieja}}]{smirnova2010shell}%
  \BibitemOpen
  \bibfield  {author} {\bibinfo {author} {\bibfnamefont {N.}~\bibnamefont
  {Smirnova}}, \bibinfo {author} {\bibfnamefont {B.}~\bibnamefont {Bally}},
  \bibinfo {author} {\bibfnamefont {K.}~\bibnamefont {Heyde}}, \bibinfo
  {author} {\bibfnamefont {F.}~\bibnamefont {Nowacki}}, and\ \bibinfo {author}
  {\bibfnamefont {K.}~\bibnamefont {Sieja}},\ }\bibfield  {title} {\bibinfo
  {title} {Shell evolution and nuclear forces},\ }\href@noop {} {\bibfield
  {journal} {\bibinfo  {journal} {Physics Letters B}\ }\textbf {\bibinfo
  {volume} {686}},\ \bibinfo {pages} {109} (\bibinfo {year}
  {2010})}\BibitemShut {NoStop}%
\bibitem [{\citenamefont {Brown}\ \emph {et~al.}(1985)\citenamefont {Brown},
  \citenamefont {Richter},\ and\ \citenamefont {Wildenthal}}]{brown1985spin}%
  \BibitemOpen
  \bibfield  {author} {\bibinfo {author} {\bibfnamefont {B.}~\bibnamefont
  {Brown}}, \bibinfo {author} {\bibfnamefont {W.}~\bibnamefont {Richter}}, and\
  \bibinfo {author} {\bibfnamefont {B.}~\bibnamefont {Wildenthal}},\ }\bibfield
   {title} {\bibinfo {title} {Spin-tensor analysis of a new empirical
  shell-model interaction for the 1s-0d shell nuclei},\ }\href@noop {}
  {\bibfield  {journal} {\bibinfo  {journal} {Journal of Physics G: Nuclear
  Physics}\ }\textbf {\bibinfo {volume} {11}},\ \bibinfo {pages} {1191}
  (\bibinfo {year} {1985})}\BibitemShut {NoStop}%
\bibitem [{\citenamefont {Osnes$~$}\ and\ \citenamefont
  {Strottman}(1992)}]{PhysRevC.45.662}%
  \BibitemOpen
  \bibfield  {author} {\bibinfo {author} {\bibfnamefont {E.}~\bibnamefont
  {Osnes$~$}}and\ \bibinfo {author} {\bibfnamefont {D.}~\bibnamefont
  {Strottman}},\ }\bibfield  {title} {\bibinfo {title} {Spin-tensor analysis of
  realistic shell model interactions},\ }\href
  {https://doi.org/10.1103/PhysRevC.45.662} {\bibfield  {journal} {\bibinfo
  {journal} {Phys. Rev. C}\ }\textbf {\bibinfo {volume} {45}},\ \bibinfo
  {pages} {662} (\bibinfo {year} {1992})}\BibitemShut {NoStop}%
\bibitem [{\citenamefont {Honma}\ \emph {et~al.}(2004)\citenamefont {Honma},
  \citenamefont {Otsuka}, \citenamefont {Brown},\ and\ \citenamefont
  {Mizusaki}}]{PhysRevC.69.034335}%
  \BibitemOpen
  \bibfield  {author} {\bibinfo {author} {\bibfnamefont {M.}~\bibnamefont
  {Honma}}, \bibinfo {author} {\bibfnamefont {T.}~\bibnamefont {Otsuka}},
  \bibinfo {author} {\bibfnamefont {B.~A.}\ \bibnamefont {Brown}}, and\
  \bibinfo {author} {\bibfnamefont {T.}~\bibnamefont {Mizusaki}},\ }\bibfield
  {title} {\bibinfo {title} {New effective interaction for $pf$-shell nuclei
  and its implications for the stability of the $n=z=28$ closed core},\ }\href
  {https://doi.org/10.1103/PhysRevC.69.034335} {\bibfield  {journal} {\bibinfo
  {journal} {Phys. Rev. C}\ }\textbf {\bibinfo {volume} {69}},\ \bibinfo
  {pages} {034335} (\bibinfo {year} {2004})}\BibitemShut {NoStop}%
\bibitem [{\citenamefont {Honma}\ \emph {et~al.}(2005)\citenamefont {Honma},
  \citenamefont {Otsuka}, \citenamefont {Brown},\ and\ \citenamefont
  {Mizusaki}}]{Honma2005}%
  \BibitemOpen
  \bibfield  {author} {\bibinfo {author} {\bibfnamefont {M.}~\bibnamefont
  {Honma}}, \bibinfo {author} {\bibfnamefont {T.}~\bibnamefont {Otsuka}},
  \bibinfo {author} {\bibfnamefont {B.~A.}\ \bibnamefont {Brown}}, and\
  \bibinfo {author} {\bibfnamefont {T.}~\bibnamefont {Mizusaki}},\ }\bibfield
  {title} {\bibinfo {title} {Shell-model description of neutron-rich pf-shell
  nuclei with a new effective interaction gxpf 1},\ }\href
  {https://doi.org/10.1140/epjad/i2005-06-032-2} {\bibfield  {journal}
  {\bibinfo  {journal} {The European Physical Journal A - Hadrons and Nuclei}\
  }\textbf {\bibinfo {volume} {25}},\ \bibinfo {pages} {499} (\bibinfo {year}
  {2005})}\BibitemShut {NoStop}%
\bibitem [{\citenamefont {Suhonen$~$}\ and\ \citenamefont
  {Civitarese}(1998)}]{suhonen1998weak}%
  \BibitemOpen
  \bibfield  {author} {\bibinfo {author} {\bibfnamefont {J.}~\bibnamefont
  {Suhonen$~$}}and\ \bibinfo {author} {\bibfnamefont {O.}~\bibnamefont
  {Civitarese}},\ }\bibfield  {title} {\bibinfo {title} {Weak-interaction and
  nuclear-structure aspects of nuclear double beta decay},\ }\href@noop {}
  {\bibfield  {journal} {\bibinfo  {journal} {Physics Reports}\ }\textbf
  {\bibinfo {volume} {300}},\ \bibinfo {pages} {123} (\bibinfo {year}
  {1998})}\BibitemShut {NoStop}%
\bibitem [{\citenamefont {Faessler$~$}\ and\ \citenamefont
  {Simkovic}(1998)}]{faessler1998double}%
  \BibitemOpen
  \bibfield  {author} {\bibinfo {author} {\bibfnamefont {A.}~\bibnamefont
  {Faessler$~$}}and\ \bibinfo {author} {\bibfnamefont {F.}~\bibnamefont
  {Simkovic}},\ }\bibfield  {title} {\bibinfo {title} {Double beta decay},\
  }\href@noop {} {\bibfield  {journal} {\bibinfo  {journal} {Journal of Physics
  G: Nuclear and Particle Physics}\ }\textbf {\bibinfo {volume} {24}},\
  \bibinfo {pages} {2139} (\bibinfo {year} {1998})}\BibitemShut {NoStop}%
\bibitem [{\citenamefont {Elliott$~$}\ and\ \citenamefont
  {Engel}(2004)}]{elliott2004double}%
  \BibitemOpen
  \bibfield  {author} {\bibinfo {author} {\bibfnamefont {S.~R.}\ \bibnamefont
  {Elliott$~$}}and\ \bibinfo {author} {\bibfnamefont {J.}~\bibnamefont
  {Engel}},\ }\bibfield  {title} {\bibinfo {title} {Double-beta decay},\
  }\href@noop {} {\bibfield  {journal} {\bibinfo  {journal} {Journal of Physics
  G: Nuclear and Particle Physics}\ }\textbf {\bibinfo {volume} {30}},\
  \bibinfo {pages} {R183} (\bibinfo {year} {2004})}\BibitemShut {NoStop}%
\bibitem [{\citenamefont {Brown}(2005)}]{brown2005lecture}%
  \BibitemOpen
  \bibfield  {author} {\bibinfo {author} {\bibfnamefont {B.~A.}\ \bibnamefont
  {Brown}},\ }\bibfield  {title} {\bibinfo {title} {Lecture notes in nuclear
  structure physics},\ }\href@noop {} {\  (\bibinfo {year} {2005})}\BibitemShut
  {NoStop}%
\bibitem [{\citenamefont {Kotila$~$}\ and\ \citenamefont
  {Iachello}(2012)}]{PhysRevC.85.034316}%
  \BibitemOpen
  \bibfield  {author} {\bibinfo {author} {\bibfnamefont {J.}~\bibnamefont
  {Kotila$~$}}and\ \bibinfo {author} {\bibfnamefont {F.}~\bibnamefont
  {Iachello}},\ }\bibfield  {title} {\bibinfo {title} {Phase-space factors for
  double-$\ensuremath{\beta}$ decay},\ }\href
  {https://doi.org/10.1103/PhysRevC.85.034316} {\bibfield  {journal} {\bibinfo
  {journal} {Phys. Rev. C}\ }\textbf {\bibinfo {volume} {85}},\ \bibinfo
  {pages} {034316} (\bibinfo {year} {2012})}\BibitemShut {NoStop}%
\bibitem [{\citenamefont {\ifmmode~\check{S}\else \v{S}\fi{}imkovic}\ \emph
  {et~al.}(2009)\citenamefont {\ifmmode~\check{S}\else \v{S}\fi{}imkovic},
  \citenamefont {Faessler}, \citenamefont {M\"uther}, \citenamefont {Rodin},\
  and\ \citenamefont {Stauf}}]{PhysRevC.79.055501}%
  \BibitemOpen
  \bibfield  {author} {\bibinfo {author} {\bibfnamefont {F.}~\bibnamefont
  {\ifmmode~\check{S}\else \v{S}\fi{}imkovic}}, \bibinfo {author}
  {\bibfnamefont {A.}~\bibnamefont {Faessler}}, \bibinfo {author}
  {\bibfnamefont {H.}~\bibnamefont {M\"uther}}, \bibinfo {author}
  {\bibfnamefont {V.}~\bibnamefont {Rodin}}, and\ \bibinfo {author}
  {\bibfnamefont {M.}~\bibnamefont {Stauf}},\ }\bibfield  {title} {\bibinfo
  {title} {$0\ensuremath{\nu}\ensuremath{\beta}\ensuremath{\beta}$-decay
  nuclear matrix elements with self-consistent short-range correlations},\
  }\href {https://doi.org/10.1103/PhysRevC.79.055501} {\bibfield  {journal}
  {\bibinfo  {journal} {Phys. Rev. C}\ }\textbf {\bibinfo {volume} {79}},\
  \bibinfo {pages} {055501} (\bibinfo {year} {2009})}\BibitemShut {NoStop}%
\bibitem [{\citenamefont {Brown$~$}\ and\ \citenamefont
  {Rae}(2014)}]{brown2014shell}%
  \BibitemOpen
  \bibfield  {author} {\bibinfo {author} {\bibfnamefont {B.}~\bibnamefont
  {Brown$~$}}and\ \bibinfo {author} {\bibfnamefont {W.}~\bibnamefont {Rae}},\
  }\bibfield  {title} {\bibinfo {title} {The shell-model code nushellx@ msu},\
  }\href@noop {} {\bibfield  {journal} {\bibinfo  {journal} {Nuclear Data
  Sheets}\ }\textbf {\bibinfo {volume} {120}},\ \bibinfo {pages} {115}
  (\bibinfo {year} {2014})}\BibitemShut {NoStop}%
\end{thebibliography}%
\end{document}